\documentclass[number]{elsarticle}

\makeatletter
\def\ps@pprintTitle{%
 \let\@oddhead\@empty
 \let\@evenhead\@empty
 \def\@oddfoot{}%
 \let\@evenfoot\@oddfoot}
\makeatother

\usepackage{pgfplots,datatool}
\usepackage{pgfplotstable,booktabs}

\usepackage{epsfig}

\usepackage{algorithm}
\usepackage[noend]{algpseudocode}
\usepackage{lineno,amsmath}
\usepackage{fixltx2e}

\usepackage[colorlinks=true]{hyperref}

\usepackage{lineno, hyperref}
\modulolinenumbers[5]

\modulolinenumbers[5] 

\pgfplotstableset{
  boldk/.style = { 
           fixed zerofill,
           postproc cell content/.append code={
            \count0=\pgfplotstablerow
            \count1=\pgfplotstablecol
            \ifnum\count0<\k 
		            \advance\count0 by1
		            \ifnum\count1<\k 
			      \advance\count1 by1
		                \pgfkeysalso{@cell content={\textbf{##1}}}%
			  \fi
            \fi
        },
 },
}

\usepackage{color, soul} 
\usepackage{amsfonts, amssymb, amsbsy, amsmath, amsthm}
\usepackage{graphics}
\usepackage{multirow, array}
\usepackage{adjustbox}
\usepackage{float}

\biboptions{authoryear}

\begin{document}

\begin{frontmatter}

\title{The Top-$K$ Tau-Path Screen for Monotone Association}
\author[h]{Srinath Sampath\corref{mycorrespondingauthor}}
\cortext[mycorrespondingauthor]{Corresponding author.}
\ead{sampath.5@osu.edu}
\author[mji]{Adriano Caloiaro}
\author[mji]{Wayne Johnson}
\author[osu]{Joseph S. Verducci}

\address[h]{Hamilton Capital Management, Columbus, Ohio, USA}
\address[mji]{Myatt \& Johnson, Inc., Miami Beach, Florida, USA}
\address[osu]{The Ohio State University, Columbus, Ohio, USA}

\begin{abstract}
A pair of variables that tend to rise and fall either together or in opposition are said to be monotonically associated.  For certain phenomena, this tendency is causally restricted to a subpopulation, as, for example, an allergic reaction to an irritant. Previously, \cite{bib.Yu:2011} devised a method of rearranging observations to test paired data to see if such an association might be present in a subpopulation.  However, the computational intensity of the method limited its application to relatively small samples of data, and the test itself only judges if association is present in some subpopulation; it does not clearly identify the subsample that came from this subpopulation, especially when the whole sample tests positive. The present paper adds a ``top-$K$'' feature (\cite{bib.Sampath:2013}) based on a multistage ranking model, that identifies a concise subsample that is likely to contain a high proportion of observations from the subpopulation in which the association is supported.  Computational improvements incorporated into this top-$K$ tau-path (TKTP) algorithm now allow the method to be extended to thousands of pairs of variables measured on sample sizes in the thousands. A description of the new algorithm along with measures of computational complexity and practical efficiency help to gauge its potential use in different settings.  Simulation studies catalog its accuracy in various settings, and an example from finance illustrates its step-by-step use.

\end{abstract}

\begin{keyword}
Kendall's tau \sep nonparametric correlation \sep ranking \sep copula \sep mixtures of distributions \sep unsupervised classification \sep computational complexity
\end{keyword}

\end{frontmatter}


\section{Introduction} \label{Introduction}

The emergence of ``big data'' has provided the opportunity for scientists to focus on key subpopulations.  These are often identified by having higher mean values on some relevant aspect that is measurable by variables in the database. Here we change the criterion to having higher correlational values.

There are many potential applications. In cancer research, the mechanisms whereby some cancers become chemoresistant are inherent in their gene networks, not all of which have been identified. One way to discover novel gene networks is to look for correlations between pairs of genes across many different types of cancer cells. The subpopulation of cancers that possess this network will match the subpopulation that supports the correlations. In marketing, a goal is to identify subpopulations that will be most responsive to a particular campaign. In finance, association between a company's earnings and a key commodity price may be restricted to certain macroeconomic conditions which are not known or directly measured. In all these cases, and more, the basic problem is to identify, as concisely as possible, a highly correlated subsample that includes most of the members of the subpopulation that supports association between the variables. This is thus a kind of unsupervised classification, with a novel criterion.

Previous work on this problem is relatively recent. \cite{bib.Yu:2011} developed the tau-path algorithm for re-sorting a sample so that the Kendall tau coefficient is (optimally) decreasing. In checking the behavior of the tail of this tau-path, they devised a test for correlation in a subsample versus the null hypothesis of independence.  A limitation of operating under this null hypothesis is that the whole population tends to test positive when the supporting subpopulation is large or when the association is strong, which is not uncommon in datasets with large sample sizes. \cite{bib.Bamattre:2015} discussed some procedures for the much more difficult problem of testing for heterogeneity of association in which the distribution under the null hypothesis may have many different forms.  A more modestly-scoped approach, suggested in \cite{bib.Sampath:2013}, assumes a positive test for overall correlation, and attempts to discover if this correlation really is supported only within a proper subpopulation. They applied a multistage model of agreement between rankings to the re-sorted sample to indicate the point where predictive information stops. This method, called the top-$K$ tau-path (TKTP), is intended to provide good coverage of the true underlying supporting population, but this has not been investigated in detail until now, mainly because the original implementation was not scalable to large datasets.

The current paper covers two issues not previously resolved. The first is extending the tau-path method to large samples. The original code was practical only up to sample sizes $n < 100$; the new code easily handles $n$ up to 10,000 on a personal computer.  This extension to large samples makes it possible to address the second issue: how to characterize the performance of TKTP beyond the simple task of testing against independence. Simulation studies demonstrate the pattern between size of selected subsample and percent of the true samples included.

Section \ref{TKTP} provides a thorough description of TKTP. Section \ref{algorithmscomputationalcomplexityandefficiency} details the design of the new algorithms, including their computational complexity and runtime efficiency. Performance characteristics of their coverage proportions appear in Section \ref{PowerAccuracy}, and a new example of predicting trends in stock prices in Section \ref{Application}. The last section of the paper summarizes all findings and discusses mathematical underpinnings.

\section{The Top-$K$ Tau-Path (TKTP) Screen} \label{TKTP}
 
This section covers Kendall's $\tau$ measure of association, its corresponding tau-path, and multistage ranking models to show how these fit together to form the TKTP method of screening for a subpopulation.

\subsection{Concordance and Kendall's $\tau$} \label{subConcordance}

A standard approach to quantifying the strength of the relationship between two variables is to enumerate the number of concordances and discordances occurring in sample pairs, and then regard the difference between these counts as a measure of association. More precisely, if $X$ and $Y$ are random variables from a joint distribution $F$, with independent observations $(X_1, Y_1)$ and $(X_2, Y_2)$, the probabilities of concordance and discordance are, respectively, 
\begin{eqnarray*}
p_c = P[(X_1 - X_2)(Y_1 - Y_2) > 0] \\
p_d = P[(X_1 - X_2)(Y_1 - Y_2) < 0],
\end{eqnarray*}
and the popular {\em Kendall's $\tau$ measure of association} is simply
\begin{eqnarray*}
\tau = p_c - p_d.
\end{eqnarray*}
Note that $\tau$ depends only on the ordinal properties of $X$ and $Y$, so that in a sample the data vectors $\mathbf X$ and $\mathbf Y$ may be replaced by their ranks with no loss of information about $\tau$.

An unbiased estimator of $\tau$, the {\em Kendall's tau coefficient}, can be computed from the random sample $\{(X_1, Y_1), (X_2, Y_2), \ldots, (X_n, Y_n)\}$ by way of the $n \times n$ concordance matrix $c = ((c_{[i,j]}))$, a tool that captures association between every sample pair in binary form as
\begin{eqnarray*}
c_{[i,j]}=
	\begin{cases}
		1, \text{ if the } (i,j)\text{th pair is concordant}\\
		-1, \text{ if the } (i,j)\text{th pair is discordant,}
	\end{cases}
\end{eqnarray*}
and from which Kendall's tau coefficient, computed as
\begin{eqnarray*}
T =  \Big( \underset{\begin{subarray}{c}
 1 \le i < j \le n
  \end{subarray}}{\sum \sum} c_{[i,j]} \Big) / \binom{n}{2},
\end{eqnarray*}
is an unbiased estimator of Kendall's $\tau$. If $A$ and $D$ are the number of concordant and discordant sample pairs, respectively, then
\begin{eqnarray*}
T = \frac{A - D}{A + D}.
\end{eqnarray*}

With this theoretical framework, Subsection \ref{subSeriation} delves into the idea of seriation, a common matrix reordering technique, and discusses the approach taken by \cite{bib.Yu:2011} to permute the concordance matrix and thus establish the sequentially maximal monotone decreasing tau-path to uncover pockets of strong association in overall moderately associated sample data.

\subsection{Seriation and the Tau-Path} \label{subSeriation}
A simple yet powerful clustering technique to discover hidden structure and underlying patterns in data involves the permutation of multidimensional data along a single dimension. In archeology, the chronological ordering of similar relics could yield new insight into the different eras based on the similarity of adjacent objects. Known as {\em seriation}, this technique finds application in a wide variety of fields of scientific inquiry. \cite{bib.Liiv:2010} provides an in-depth history of this technique and its use in many disciplines. In unsupervised learning, seriation often manifests in the form of matrix reordering, where multidimensional matrices are permuted based on the magnitude of a particular characteristic, potentially leading to insights into underlying associations in the data. 

In the context of Kendall's $\tau$, \cite{bib.Yu:2011} permuted the sample concordance matrix in a very specific manner described in the next section. By means of either of two backward conditional search (BCS) algorithms, the Fast Backward Conditional Search (FastBCS) and the Full Backward Conditional Search (FullBCS) algorithms, they selected increasing subsets of the sample matrix in such a way that the associated tau coefficients become monotonically decreasing; the tau coefficient corresponding to the $2 \times 2$ subset is therefore at least as large as the one corresponding to the $3 \times 3$ subset, and so on. The resulting ordering of the sample matrix is referred to as the tau-path.

Using the tau-path statistic, \cite{bib.Yu:2011} also propose a test for independence between the two variables versus the alternative that the variables are correlated only in a subpopulation. By simulating a large number of independent samples of size $n$ from independent $X$ and $Y$ populations, establishing the tau-path for each sample, and then 
calculating the percentiles of Kendall's tau coefficient at each step along the tau-path, and picking a fixed percentile to serve as the boundary at each step; for any fixed percentile $q$, the boundary points $\{q_i|\  i = 2, \ldots, n \}$ are calibrated so that no more than a proportion $\alpha$ of the random tau-paths exceed the boundaries. Note that the same simulation may be achieved by sampling pairs of independent permutations of $\{1,\ldots,n\}$.

The tau-path approach thus provides an ordering of the sample bivariate data along the path of strongest association. So far the focus of this subsection has been on uncovering positive associations between the two variables; should a negative association between the variables be sought, this is readily accomplished by multiplying the $Y$'s by $-1$ before performing the tau-path analysis. 

Another aspect of the ordered data that is worth computing is the {\em stopping stage} or {\em endpoint of association}; the stage in the tau-path ordered data where association ends and randomness sets in. Especially in these times when large datasets are readily available, and speed of computation and immediacy of results are important considerations, the need to prune the full dataset down to the appropriately-sized subset for further analysis is critical. The key notion here is that as one progresses down the tau-path from most- to least-associated sample pairs, the two data elements in any pair should provide largely consistent information about their strength of association. When adjacent data pairs routinely provide inconsistent information about their strength of association, it is possible that the level of association has deteriorated and randomness has taken over. In the context of the tau-path ordering, multistage ranking models provide an appropriate framework for extracting the endpoint of agreement.

\subsection{The Multistage Ranking Model and the TKTP Stopping Rule} \label{subMultistage}

The primary challenge with traditional models that assess the degree of association between two ranked lists is their mathematical complexity. Whereas a multinomial distribution provides a full characterization of the population, the large number of ensuing parameters leads to intractable likelihood equations, even when the number of objects being ranked is small. \cite{bib.Fligner:1988} proposed a family of forward multistage ranking models that vastly reduce the computational effort and also provide clear insights into the underlying processes. The multistage ranking framework is outlined here and the TKTP stopping rule then developed. 

Given a group of $n$ objects, the relative value of the objects to an assessor---whether qualitative or numerical---can be expressed by the assignment of ranks to the objects. Two representations of the assessor's preferences are popular; the first is the {\em ranking} or {\em permutation}, a vector of length $n$ that gives, for each object on the original list, the assessor's respective rank, namely the quantity (1 + the number of other objects that the assessor  considers superior). This representation of the ranked preferences is denoted
\begin{eqnarray*}
\pi =[\pi(1), \ldots, \pi(n)].
\end{eqnarray*}

The second representation, an {\em ordering} or {\em inverse permutation} of the $n$ objects, is also a vector of length $n$, showing the labels of the $n$ objects set in ranked order. Thus the ordering or inverse permutation corresponding to the representation $\pi$ above is
\begin{eqnarray*}
\pi^{-1}(j) = i~{\text {if}}~\pi(i) = j,~i = 1, \ldots, n,~j = 1, \ldots, n.
\end{eqnarray*}

Our interest lies in measuring the extent of agreement between the preferences of two assessors who independently rank the $n$ objects, and in determining the last stage where agreement still exists and random noise follows, the {\em stopping stage} or {\em endpoint of agreement} between the two assessors. This is achieved by anchoring one assessor's ranking of the objects as the {\em reference ranking} or {\em ground truth}, and then examining the stage-wise departures of the second assessor's ranks from the corresponding reference ranks; the latter is the {\em generated} or {\em observed ranking} $\pi$. Following the approach given by \cite{bib.Fligner:1988}, the computations for the first stage and a generic stage $j$ are given below. The stages are assumed to be independent, and penalties and truncated geometric probabilities are assigned to the second assessor's ranks commensurate with their stage-wise deviations from the first assessor's ranks as follows:

Stage 1: Since all $n$ objects are available, the second assessor selects the $(1 + v)$th best object {\em overall}, as specified by  $\pi^{-1}$, and incurs the penalty $V_1 = v$ with probability
\begin{eqnarray*} 
P(V_1 = v) = \Bigg(\frac{1-r_1}{1-r_1^{n}}\Bigg) r_1^v,~v = 0,\ldots,n-1,~0<r_1<1. 
\end{eqnarray*}

Stage $j~(j= 2, \ldots, n-1)$: For each subsequent stage $j$,~$n-j+1$~objects are still available. Here the second assessor selects the $(1 + v)$th best {\em available}~object, as specified by $\pi^{-1}$,~and incurs a penalty $V_j = v$ with probability
\begin{eqnarray*} 
P(V_j = v) = \Bigg(\frac{1-r_j}{1-r_j^{n-j+1}}\Bigg) r_j^v,~v = 0,\ldots,n-j,~0<r_j<1. 
\end{eqnarray*}

The vector $\{V_1, \ldots, V_{n-1}\}$ thus captures the stage-wise deviations between the two sets of ranks, and is referred to as the {\em discordance} or {\em penalty vector} between the ranking schemes. $\{r_j\}$ too represents the stage-wise disagreement between the two assessors. Rather than focus on $r_j$, the transformation 
\begin{eqnarray*}
\theta_j = -\log r_j,~j = 1, \ldots, n-1
\end{eqnarray*}
is used, and a higher estimated $\theta_j$ reveals a closer association between the assessors' ranks for object $j$.

It remains to determine the stopping stage or endpoint $K$, the last stage in the list of objects where the second assessor's rank still shows some form of agreement with that of the first. Beyond this stage, the second assessor posts random stage-wise ranks relative to the first, and the previous association between the assessors' ranks has faded. Subsection 3.1 of \cite{bib.Sampath:2013} provides the moving average maximum likelihood estimator (MAMLE) for the $r_j$, calculated iteratively over overlapping backward-looking windows of stages $i$, $j-w+1 \le i \le j$ of fixed width $w$, till all the stages are exhausted. The resulting curve $\{\hat{r}_j\}$ and its inverted counterpart $\{\hat{\theta}_j\}$ are locally smooth estimators of the stage-wise agreement between the two assessors.

The rejection region for the MAMLE, which enables the stopping rule, is now straightforward. Briefly, a large number of simulations are generated from the multistage model under the assumption that all the $\theta_j$'s are 0, and stage-wise $\hat{\theta}_j$'s are computed using the MAMLE method. For each stage $j$, the $(1-\alpha)$th quantile $q(j)$ is computed. An estimate of the endpoint $K$ is given by
\begin{multline}
\hat{K} = \text{the earliest stage at which } \hat{\theta}_{\hat{K}+w} > q(\hat{K}), \text{ and } \hat{\theta}_j > q(j) \text{ for at most }\\ \alpha \text{ percent } \text{of the remaining } j > \hat{K} + w. \label{StoppingRule}
\end{multline}

Additional diagnostics and guidelines on using the stopping rule are provided in Subsection 3.2 of \cite{bib.Sampath:2013}. An alternative approach, based on a data-analytic method, is covered by \cite{bib.Hall:2012}.

The two techniques discussed above accomplish very distinct goals; the tau-path algorithm organizes bivariate data along the path from strongest to weakest association, whereas the top-$K$ MAMLE approach detects the endpoint of association between two ranked lists. Synthesis of the algorithms is accomplished by using the tau-path ordering from the first algorithm as the ordering of the stages of the second. 

Here the framework for the multistage model is that the pair $(X, Y)$ of variables plays the role of a rater, ranking the observations $\{1, \ldots, n\}$ according to the tau-path ordering.  The actual values of Kendall's tau coefficient $T_k$ over the first $k$ stages along the tau-path are used to infer the strength of association at each stage. Choosing the stages according to the tau-path guarantees that the estimated $\theta_j$ will be decreasing. This happens because the number of new discordances introduced at each tau-path stage must be non-decreasing. In this context, stopping rule (\ref{StoppingRule}) identifies observations in the remaining stages to have $(X, Y)$ association no greater than noise. Since the ordered data is used as the input into the top-$K$ algorithm, the computed endpoint gives the stopping stage where the strongly associated subsample ends. This subsample may be studied further to try to infer the underlying subpopulation.  Section \ref{Application} provides an example where several pairs of variables are used to estimate a common subsample whose import may be identified by the behavior of an explanatory variable on this subsample.

\section{Algorithms, Computational Complexity and Efficiency of Implementation}
\label{algorithmscomputationalcomplexityandefficiency}

The algorithms that underlie the TKTP method were developed to screen out observations for which a pair $(X, Y)$ of variables have association no greater than noise. The remaining observations are then most likely to represent a subpopulation that supports strong association. For large datasets this computational task can be daunting. For example, in the analysis of microarrays used in molecular biology, gene chip platforms support increasing numbers of gene probes that can be attached to a microarray. In large-scale toxicity studies, hundreds of microarrays may be used to measure the effects of chemical compounds at various dosage levels over time. In a study measuring the expression levels of 30,000 genes responding to 3,600 chemical compound treatments, a dataset of 3,600 observations and 30,000 variables would be generated. In screening for relationships between pairs of genes, $30,000 \choose 2$ or 449,985,000 comparisons would be made. For each gene-gene pair identified, up to $n =$ 3,600 observations could be combined into $ {\sum_{k = 2}^{n} {n \choose k}/2} $ sets from sizes 2 to \emph{n} in the search for associated subsets.

For large datasets, it is infeasible to examine computationally all pairs of variables and the population subsets within each pair. Critical to the design of the top-$K$ and tau-path algorithms was an understanding of their order of growth. The variety of software languages and hardware architectures we considered required care not to bias the specification of the algorithms toward a particular implementation context since programming languages, compilers, processor architectures, and memory hierarchies all affect the performance of any implementation. 

\begin{figure}[ht] 
\centering{
\includegraphics[scale=0.7]{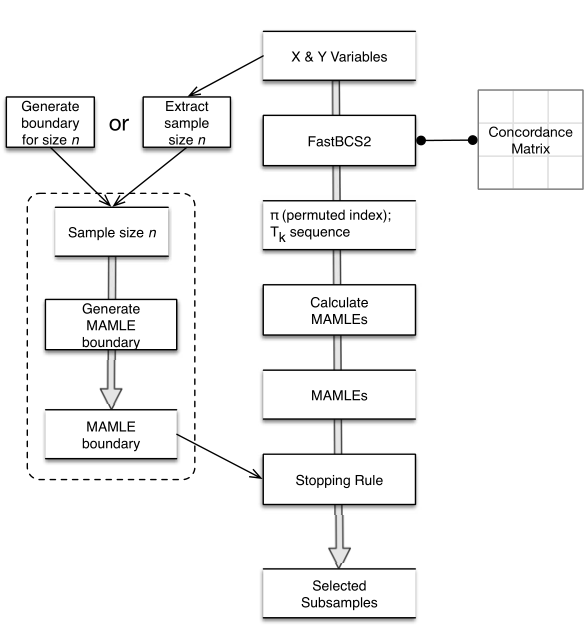}} 
\caption{The TKTP algorithm.}
\label{FigTKTPAlgorithm}
\end{figure}  

The TKTP algorithm is the top-level algorithm. A flowchart for this algorithm appears in [Figure \ref{FigTKTPAlgorithm}]. In Subsection \ref{overviewofalgorithms} we provide an overview of the major steps of the TKTP algorithm and the functions it invokes. Since FastBCS and its optimized successor FastBCS2 are critical to the runtime performance of TKTP, these algorithms are described in detail in Subsection \ref{fastbcsandfastbcs2} along with a simple example showing how they work. In Subsection \ref{computationalcomplexity} we analyze the computational complexity of the FastBCS* algorithms, and in Subsection \ref{efficiencyandimplementationstrategies} describe their efficiency and the various strategies we explored for implementing them. The pseudocode used to describe the algorithms mostly follows the conventions specified in \cite{bib.Cormen:2009}.

\subsection{Overview of Algorithms}
\label{overviewofalgorithms}

\subsubsection{The TKTP Wrapper} \label{tktp}

The input to the TKTP algorithm shown in [Algorithm \ref{algorithm-tktp}] is a random sample of observations for variables $X$ and $Y$ from a bivariate distribution. It invokes four major functions to find a subset of strongly associated observations within this pair: FastBCS, TaupathMAMLE, GenerateRejectBoundary, and StoppingPoint:

\begin{algorithm}[H] 
\caption{TKTP($X, Y$)}
\label{algorithm-tktp}
\begin{algorithmic}[1]

    \Statex \textbf{Input:} A random sample $S$ of $n$ pairs $(X_1, Y_1),$ $\ldots,$ $(X_n, Y_n)$  from a continuous bivariate population.
    \Statex \textbf{Output:} A possibly empty subset $s$ of observations in $S$ in which $X_s$ and $Y_s$ are strongly associated.
    \Require $length(X) = length(Y)$

    \State $WINDOW \gets 5$
    \Statex $SIGLVL \gets 0.05$
    \Statex $NSIM \gets 10000$
    \Statex // Find the permuted order of the observation pairs and the tau-path for $X$ and $Y$ using FastBCS*.
    \State $pi, taupath \gets FastBCS2(X, Y)$
    \Statex // Get moving average maximum likelihood estimators (MAMLEs) for $X$ and $Y$.
    \State $\hat{\theta} \gets TaupathMAMLE(taupath, WINDOW)$
    \Statex // Simulate the MAMLE boundary at the $(1-\alpha)$th quantile of the null hypothesis.
    \State $boundary \gets GenerateRejectBoundary(n, WINDOW, NSIM, SIGLVL)$ 
    \Statex // Estimate the stopping point $\hat{K}$.
    \State $\hat{K} \gets StoppingPoint(\hat{\theta}, boundary, SIGLVL)$
    \State
    \Return $\{pi[j]\ |\ [ j \geq \hat{K}] \wedge [\hat{\theta}_j > q(j)]\}$, where $q(j)$ is the $(1 - \alpha)$th quantile of $\hat{\theta}_j$

\end{algorithmic}
\end{algorithm}

\textbf{Line 2.} The FastBCS algorithm shown in [Algorithm \ref{algorithm-fastbcs}] orders the pairs from the strongest to least associated and generates the tau-path of Kendall's tau coefficients that measure the strength of the association for each nested subset from size 2 to $n$. 

\begin{algorithm}
\caption{FastBCS($X$, $Y$)} 
\label{algorithm-fastbcs}
\begin{algorithmic}[1]
\Statex \textbf{Input:} A random sample of  $n$ pairs $(X_1, Y_1), \ldots, (X_n, Y_n)$  from a continuous bivariate population.
\Statex \textbf{Output:} The final permutation $pi$ and the tau-path statistic derived from the concordance matrix constructed from $pi$.
\Require  $n$ = $length(X) = length(Y)$; $n > 1$ 
\Statex // The function $indexOf(a, v)$ returns the index of the value $v$ in array $a$.
\State $C \gets$ the concordance matrix for $(X_1, Y_1), \ldots, (X_n, Y_n)$ 
\Statex $pi \gets [1 \ldots n]$ 
\Statex $i \gets n$ 
\Statex $tie[1 \ldots n] \gets \emptyset$ 

\Repeat
    \Comment backward conditional search
    \For{$j \gets 1 \ldots i$}
    \Comment backward elimination
        \State $colsum[j] \gets \sum_{u=1}^i C_{[pi[u], pi[j]]}$
    \EndFor

    \State $minsum \gets$ minimum value in $colsum$
    \State $ties\_i \gets \{pi[l], l \gets 1 \ldots i | colsum[pi[l]] = minsum \}$

    \If{$|ties\_i| > 1$}
        \State $tie[i] \gets ties\_i$
        \State $r\_tie \gets$ randomly select a member of $ties\_i$
        \State $l \gets indexOf(pi, r\_tie)$
        \State transpose $pi[i]$ and $pi[l]$
    \Else
        \State $l \gets indexOf(pi, ties\_i[1])$
        \State transpose $pi[i]$ and $pi[l]$
    \EndIf

    \For {$k \gets$ $n$  \textbf{downto} $(i+1)$}
        \Comment tie logic
        \If {$pi[i] \in tie[k]$}
            \State $qi \gets [\sum_{u=1}^i C_{[pi[u],pi[i]]}, \sum_{u=1}^{i+1} C_{[pi[u],pi[i]]}, \ldots ,  \sum_{u=1}^k C_{[pi[u],pi[i]]}]$
            \State $qk \gets [\sum_{u=1}^i C_{[pi[u],pi[k]]}, \sum_{u=1}^{i+1} C_{[pi[u],pi[k]]}, \ldots , \sum_{u=1}^k C_{[pi[u],pi[k]]}]$
            \If { $All(qk[u] \geq qi[u]) and Any(qk[u] > qi[u]), u=1 \ldots k-i+1$ }
                \State transpose $pi[i]$ and $pi[k]$
                \State $i \gets k - 1$
                \Comment forward step
                \State $tie[m] \gets [], \forall m \leq k$
                \State GOTO \textbf{repeat}
            \EndIf
        \EndIf
    \EndFor

    \State $i \gets i - 1$
    \Comment backward step
\Until $(\sum_{j=1}^i \sum_{k=1}^i C_{[pi[k],pi[j]])}) = (i * (i - 1))$

\State $k \gets i$
\State $T[2] \gets \dots \gets T[k] \gets 1$
\State
\Return The final permutation $pi$ and the tau-path statistic \{$T[2] \ldots T[n]$\} 

\end{algorithmic}
\end{algorithm}

\textbf{Line 3.} The tau-path and window are passed as arguments to the Taupath\-MAMLE algorithm shown in [Algorithm \ref{algorithm-tp-mle}] to generate a MAMLE curve for the $X$ and $Y$ pair. The tau-path is used in the penalty function to calculate the MAMLE [4] curve. The window width determines the smoothness of the curve [6--8]. This algorithm is used to generate the reference ranking of the first assessor's choices for the $X$ and $Y$ pair of variables, and to simulate the second assessor's choices under the null hypothesis.

\begin{algorithm}
\caption{TaupathMAMLE(tauPath, window)}
\label{algorithm-tp-mle}
\begin{algorithmic}[1]

\Statex \textbf{Input:} 
\Statex - The tau-path of a random sample of  $n$ pairs $(X_1, Y_1), \ldots , (X_n, Y_n)$   from a continuous bivariate population.
\Statex - The size of the window.
\Statex \textbf{Output:} An array of MAMLEs. 
\Require   $n = length(tauPath)$; $n > 1$   

\State $ma.theta \gets meanDiff \gets [ \  ]$
\State \emph{diffs} $\gets [0]$
\For{$i \gets 2 \ldots n$}  
    \State $totaldiscord[i-1] \gets \left( (1 - tauPath[i])/2) * {i \choose 2} \right)$
    \State \emph{diffs}$[i] \gets (totaldiscord[i]  - totaldiscord[i - 1])$
\EndFor

\For{$i \gets 0 \ldots (n-window-1)$}   
    \State $meanDiff \gets diffs[(i+1) \ldots (i + window)]$
    \State $ma.theta[i + window + 1] \gets theta.scale(meanDiffs, i, window, n+1)$   
\EndFor

\State \Return ma.theta

\end{algorithmic}
\end{algorithm}

\textbf{Line 4.} To determine the rejection region under the null hypothesis, the GenerateRejectBoundary function (not shown) simulates the second assessor's choices under the null hypothesis by generating the MAMLE curves for a pair of random variables of size $n$   for the specified number of iterations and invoking TaupathMAMLE on each pair. From these MAMLE curves, the stage-wise $ (1-\alpha) $th quantiles are calculated and the boundary representing the edge of the rejection region is returned.

\textbf{Line 5.} The StoppingPoint algorithm shown in [Algorithm \ref{algorithm-stopping-point}] estimates the \emph{stopping stage} or \emph{endpoint K} that indexes the end of agreement between the reference ($XY$ tau-path) and generated (rejection boundary) rankings, by $\hat{K}$.

\begin{algorithm}
\caption{StoppingPoint(xyMamles, boundary, significanceLevel)}
\label{algorithm-stopping-point}

\begin{algorithmic}[1]
\Statex \textbf{Input:} 
\Statex - The MAMLEs generated for the $X$ and $Y$ variables.
\Statex - The rejection boundary under the null hypothesis.
\Statex - The significance level of the rejection boundary.
\Statex \textbf{Output:} The estimated value of the stopping point $K$ or 0 if none was found.
        
\Require   $n$ = $length(xyMamles) = length(boundary)$; $n > 1$   
    \State $exceed \gets [ \thinspace ]$
    \Statex $j \gets 1$
    \For{$i$ in $1 \ldots n$}  
        \If{$mamles[i] > boundary[i]$}
            \State $exceed[j] \gets i$
            \State $j \gets j + 1$
        \EndIf
    \EndFor
    \State $numExceed \gets length(exceed)$
    \State $sort(exceed)$
    \For{i in $1 \ldots length(exceed)$}
        \State $tail \gets n - exceed[i]$
        \State $left \gets numExceed - i$
        \If{$left \leq significanceLevel * tail$}
            \State \Return candidate[$i$]
        \EndIf  
    \EndFor
    \State \Return 0
  \end{algorithmic}
\end{algorithm}

\subsubsection{FastBCS and FastBCS2}
\label{fastbcsandfastbcs2}

The Kendall's tau coefficient of the concordance matrix (\cite{bib.Yu:2011}; p. 101) of a pair of variables is defined as
\begin{equation}
T = \frac{A - D}{{n \choose 2}} \label{KendallTau}
\end{equation}
where $n \choose 2$ is the number of distinct pairs of observations in the sample, and $A$ and $D$ are the number of concordant and discordant pairs of observations, respectively. The tau-path is a monotonically decreasing sequence of Kendall tau coefficients. The FastBCS algorithm (\cite{bib.Yu:2011}; p. 103) generates the tau-path of a pair of statistical variables $X$ and $Y$ of a sample of observations. We have restated the algorithm in [Algorithm \ref{algorithm-fastbcs}] and include a simple example to help explain it. All numbers in brackets below refer to line numbers in the FastBCS algorithm.

The backward conditional search [2--25] consists of two main segments known as \emph{backward elimination} [3--14] and \emph{tie logic} [15--23]. The FastBCS incrementally constructs the permutation sequence $pi$ backward from the set of all observations $S_n$ at $i = n$   to a subset of two observations $S_2$ at $i = 2$. The tau-path is constructed using the permuted order of $pi$ after the main \emph{repeat} loop exits. Each iteration \emph{i} is called a \emph{stage}. 

Prior to the search, the main data structures used during the operations are initialized [1]. The concordance matrix \emph{C} of the set of observations for $i = n$   is calculated using the natural ordering of the variables $X$ and $Y$ provided as input to the algorithm. The values of the permuted index \emph{pi} used to track the reordering of the observations at each stage are initialized to the natural ordering of $X$ and $Y$. At each stage \emph{i}, an $i \times i$ permuted matrix will be derived from \emph{C} using \emph{pi}. This is shown in the figures below as $C_i$. The row and column headings of the permuted concordance matrix are the values indexed by \emph{pi} and refer to the observation IDs shown in the headings of \emph{C}. The \emph{tie} array maintains a \emph{tieset} for each stage. Each element of the \emph{tie} array is set to empty tiesets.

The search begins with backward elimination. The subset $S_i$ of observations $pi[1 \ldots i]$ for any stage \emph{i} is fixed. The goal of backward elimination is to find and eliminate the observation in the subset $pi[1 \ldots i]$ that contributes least to Kendall's tau coefficient---or increases most the value of \emph{D} in Equation (\ref{KendallTau})---of the remaining subset of observations $S_{i - 1}$. Elimination is done by transposing this observation with the observation at $pi[i]$. The result of the transposition guarantees that $T_{i-1} \geq T_i$ in this stage. If $pi[i]$, the observation represented by the \emph{i}th stage, is not a member of any prior stages' tiesets, a backward step is taken by setting $i$ to $i - 1$ [24]. These steps are repeated until either Kendall's tau coefficient for stage \emph{i} becomes 1, or $ i = 2 $ [25]. 

A tie occurs at stage $i$ if more than one observation could be eliminated from $S_i$ [7]. The tied observations form a tieset that is associated with stage \emph{i} [8]. The observations in tiesets may be reexamined in the tie logic of later stages [16]. To complete the backward elimination at stage \emph{i} in the presence of ties, one observation is arbitrarily chosen and eliminated from $S_i$ [9]. Because the tau-path is constructed in reverse order, the algorithm cannot determine the effect of future rearrangements on local maximal monotonicity at the time a selection from the tieset is made. In subsequent stages $ k < i $, the tie logic will reexamine the choice made in stage \emph{i} [17--18] and take a forward step [21], if necessary, to ensure monotonicity in the tau-path. To convey an intuitive understanding for how the algorithm works, we walk through a simple example.

\begin{figure}[ht]
\renewcommand{\arraystretch}{1.2}
\begin{minipage}[c][1\totalheight][t]{1\linewidth}%
\begin{center}
\begin{minipage}[c][1\totalheight][t]{0.5\linewidth}%
\textbf{Initialization}
\begin{quote}
$C \gets$ concordance matrix of $X$, $Y$ \\
$pi \gets [1, 2, 3, 4, 5]$ \\
$i \gets 5$ \\
$tie[1 \ldots 5] \gets \emptyset$\end{quote}
\end{minipage}
\begin{tabular}{c|rrrrr|}
\multicolumn{1}{c|}{\textbf{$C$}} & \textbf{1} & \textbf{2} & \textbf{3} & \textbf{4} & \multicolumn{1}{r}{\textbf{5}}\tabularnewline
\hline 
\textbf{1} & 0 & -1 & -1 & 1 & -1\tabularnewline
\textbf{2} & -1 & 0 & -1 & 1 & -1\tabularnewline
\textbf{3} & -1 & -1 & 0 & -1 & 1\tabularnewline
\textbf{4} & 1 & 1 & -1 & 0 & -1\tabularnewline
\textbf{5} & -1 & -1 & 1 & -1 & 0\tabularnewline
\cline{2-6} 
\end{tabular}
\par\end{center}
\end{minipage}

\caption{Example: the initialization of the concordance matrix $C$ calculated for a pair of  variables $X$ and $Y$ with five observations.}\label{figureInitializationConcordance}

\end{figure}

\emph{Example.} $ \; $ Suppose we wish to generate a tau-path of the variables $X = [1, 2, 4, 3, 5]$ and $Y = [4, 3, 1, 5, 2]$ of a sample of five observations. FastBCS is invoked with $X$ and $Y$ as arguments. The initialization [1] is shown in [Figure \ref{figureInitializationConcordance}]. It initializes the concordance matrix \emph{C} based on the natural ordering of the pairs of observations in $X$ and $Y$. The array \emph{pi} is set to the natural ordering 1, \ldots, 5. The current stage \emph{i} is set to 5. Finally, each element of array \emph{tie} which preserves the ties found at a particular stage $i$ is initialized to the empty set. [Algorithm \ref{algorithm-fastbcs}] begins the backward conditional search [2] and the first two stages, stages 5 and 4, are shown in [Figure \ref{figureMainSteps5}].

\begin{figure}[ht]

\begin{minipage}[c][1\totalheight][t]{1\columnwidth}
\begin{center}
\begin{minipage}[c][1\totalheight][t]{0.5\columnwidth}
\textbf{Stage 5}

Backward Elimination
\begin{quote}
Calculate the column sums. \\
Create a list of ties: \\$ties\_i \gets [1, 2, 3, 5]$. \\
Transpose $pi[5]$ and $pi[1].$
\end{quote}
Tie Logic
\begin{quote}
There are no tiesets to examine in stage 5.
\end{quote}
backward step: $i \gets 4$
\end{minipage}\textbf{}
\begin{tabular}{c|rrrrr|}
\textbf{$C_{5}$} & \textbf{1} & \textbf{2} & \textbf{3} & \textbf{4} & \multicolumn{1}{r}{\textbf{5}}\tabularnewline
\hline 
\textbf{1} & 0 & -1 & -1 & 1 & -1\tabularnewline
\textbf{2} & -1 & 0 & -1 & 1 & -1\tabularnewline
\textbf{3} & -1 & -1 & 0 & -1 & 1\tabularnewline
\textbf{4} & 1 & 1 & -1 & 0 & -1\tabularnewline
\textbf{5} & -1 & -1 & 1 & -1 & 0\tabularnewline
\hline 
\hline 
$colsums$ & -2 & -2 & -2 & 0 & \multicolumn{1}{r}{-2}\tabularnewline
$ties\_i$ & 1 & 2 & 3 &  & \multicolumn{1}{r}{5}\tabularnewline
\hline 
\end{tabular}
\par\end{center}

\begin{center}
\begin{minipage}[c][1\totalheight][t]{0.5\columnwidth}
\textbf{Stage 4}

Backward Elimination
\begin{quote}
Calculate the column sums. \\
Create a list of ties: \\$ties\_i \gets [5, 2, 3, 4]$. \\
Transpose $pi[4]$ and $pi[1]$.
\end{quote}
Tie Logic
\begin{quote}
$pi[4]=4$ is not a member of any tieset for $k>4$.
\end{quote}
backward step: $i \gets 3$
\end{minipage}\textbf{}
\begin{tabular}{c|rrrr|r}
\textbf{$C_{4}$} & \textbf{5} & \textbf{2} & \textbf{3} & \multicolumn{1}{r}{\textbf{4}} & 1\tabularnewline
\hline 
\textbf{5} & 0 & -1 & 1 & -1 & -1\tabularnewline
\textbf{2} & -1 & 0 & -1 & 1 & -1\tabularnewline
\textbf{3} & 1 & -1 & 0 & -1 & -1\tabularnewline
\textbf{4} & -1 & 1 & -1 & 0 & 1\tabularnewline
\cline{2-5} 
1 & -1 & -1 & -1 & \multicolumn{1}{r}{1} & 0\tabularnewline
\hline 
\hline 
$colsums$ & -1 & -1 & -1 & \multicolumn{1}{r}{-1} & \tabularnewline
$ties\_i$ & 5 & 2 & 3 & \multicolumn{1}{r}{4} & \tabularnewline
\hline 
\end{tabular}
\par\end{center}
\end{minipage}

\caption{Example: the main steps taken in stage 5 resulting in a backward step to stage 4.}\label{figureMainSteps5}

\end{figure}

All five observations from the sample are in subset $S_5$. To find the observation to eliminate from $S_5$, backward elimination begins by calculating the column sums of each of the observations from 1 to 5 of the permuted concordance matrix $C_5$ [3--4]. Because no transpositions have taken place, $C_5$ is identical to \emph{C}. Four of the columns of $C_5$ sum to values of --2, so we generate the tieset of observations $\{1, 2, 3, 5\}$ [6] and save it in $tie[5]$ [8]. Although any of these four observations could have been randomly chosen for elimination, throughout this example we will always choose the first observation of the tieset. In stage 5, by transposing $pi[1]$ with $pi[5]$, the observation that contributes least to the Kendall's tau coefficient of $S_4$, \emph{pi} is reordered so that the observations from $pi[1]$ to $pi[4]$ will generate $T_4 \geq T_5$. No operations are performed in the tie logic since there is no $k > i$. We set \emph{i} to 4 [24] and proceed in similar fashion at stage 4 with $pi = [5, 2, 3, 4, 1]$. Our interest in stage 4 is only in the first four elements of \emph{pi} which constitute the subset $S_4$ and the concordance matrix $C_4$. Four columns are summed and the tieset $\{5, 2, 3, 4\}$ is saved in $tie[4]$. Since item 4 is not a member of any tieset of $k>i$, there are no prior tiesets to reexamine, so items 4 and 5 are transposed and $i$ is set to 3. In both stages 4 and 5, only backward steps have been taken.

Stage 3 is shown in [Figure \ref{figureMainSteps3}]. The steps in backward elimination find the sole observation 3 to contribute the least. No transposition is required. To ensure that $pi[3] = 3$ locally optimizes the tau-path, the tie logic [15--23] reexamines all tiesets from \emph{k} down to $i + 1$ in which 3 is a member. Since observation 3 is a member of the tieset from stage 5, the processing continues by attempting to determine whether the tau coefficient at stage 3 could have been improved had the observation at $pi[k=5]$ been chosen instead. The calculations are shown in the tie logic of [Figure \ref{figureMainSteps3}]. In effect, the algorithm compares the cumulative sums of stage 3 of two different concordance matrices shown as $C_3$ and $C_3^\prime$ in [Figure \ref{figureMainSteps3}(a) and \ref{figureMainSteps3}(b)], respectively. $C_3$ represents the permuted order from the choice that was made in stage 5. $C_3^\prime$ represents the concordance matrix that would have resulted from choosing the observation $pi[k=5]$, and is constructed by a transposition of the observations at $pi[3]$ and $pi[5]$. The cumulative sum $qi$ is calculated from $C_3$ and $qk$ from $C_3^\prime$ [17-18]. Two tests must pass [19] if $C_3^\prime$ is to be considered the better choice: All of $qk_{u=3}^5 \geq qi_{u=3}^5$ and Any $qk_{u=3}^5 > qi_{u=3}^5$. Both tests succeed and the algorithm continues with $C_3^\prime$ as shown in [Figure \ref{figureMainSteps3}(c)].

\begin{figure}[ht!]
\textbf{Stage 3}

\begin{tabular}[t]{>{\raggedright}m{0.5\columnwidth}r}
\phantom{}

\phantom{}

Backward Elimination
\begin{quote}
Calculate column sums.\\
No ties to save.
\end{quote}
Tie Logic
\begin{quote}
$pi[3]=3$ is a member of stage 5 tieset.\\
Determine locally optimal observation.\end{quote}
 & \textbf{}%
\begin{tabular}{r|rrrrr}
\textbf{$C_{3}$} & \textbf{4} & \textbf{2} & \textbf{3} & 5 & 1\tabularnewline
\hline 
\textbf{4} & 0 & 1 & \multicolumn{1}{r|}{-1} & -1 & 1\tabularnewline
\textbf{2} & 1 & 0 & \multicolumn{1}{r|}{-1} & -1 & -1\tabularnewline
\textbf{3} & -1 & -1 & \multicolumn{1}{r|}{0} & 1 & -1\tabularnewline
\cline{2-4} 
5 & -1 & -1 & 1 & 0 & -1\tabularnewline
1 & 1 & -1 & -1 & -1 & 0\tabularnewline
\hline 
\hline 
colsums & 0 & 0 & -2 &  & \tabularnewline
ties\_i &  &  &  &  & \tabularnewline
\hline 
\end{tabular}\tabularnewline
 & \tabularnewline
(a) Calculate the cumulative sum $qi$ of $pi[i=3]$ & %
\begin{tabular}{r|rrrrr}
\multicolumn{1}{r}{} &  &  & $i$ &  & $k$\tabularnewline
\textbf{$C_{3}$ } & 4 & 2 & \textbf{3} & 5 & \textbf{1}\tabularnewline
\hline 
4 & 0 & 1 & -1 & -1 & 1\tabularnewline
2 & 1 & 0 & -1 & -1 & -1\tabularnewline
\cline{4-4} 
\textbf{3} & -1 & \multicolumn{1}{r|}{-1} & \multicolumn{1}{r|}{0} & 1 & -1\tabularnewline
\textbf{5} & -1 & \multicolumn{1}{r|}{-1} & \multicolumn{1}{r|}{1} & 0 & -1\tabularnewline
\textbf{1} & 1 & \multicolumn{1}{r|}{-1} & \multicolumn{1}{r|}{-1} & -1 & 0\tabularnewline
\hline 
\end{tabular}\tabularnewline
 & \tabularnewline
 & \tabularnewline
(b) Calculate the cumulative sum $qk$ of $pi[k=3]$ & %
\begin{tabular}{r|ccrcr}
\multicolumn{1}{r}{} &  &  & \textit{k} &  & \textit{i}\tabularnewline
$C_{3}^{\prime}$ & 4 & 2 & \textbf{1} & 5 & \textbf{3}\tabularnewline
\hline 
4 & 0 & 1 & 1 & -1 & -1\tabularnewline
2 & 1 & 0 & -1 & -1 & -1\tabularnewline
\cline{4-4} 
\textbf{1} & -1 & \multicolumn{1}{c|}{-1} & \multicolumn{1}{r|}{0} & 1 & -1\tabularnewline
\textbf{5} & -1 & \multicolumn{1}{c|}{-1} & \multicolumn{1}{r|}{-1} & 0 & 1\tabularnewline
\textbf{3} & 1 & \multicolumn{1}{c|}{-1} & \multicolumn{1}{r|}{-1} & -1 & 0\tabularnewline
\hline 
\end{tabular}\tabularnewline
 & \tabularnewline
 & \tabularnewline
(c) $All(qk \ge qi)$ AND $Any(qk > qi)$ test passes. & %
\begin{tabular}{r|r}
$qi$ & \textit{qk}\tabularnewline
\hline 
-2 & 0\tabularnewline
-1 & -1\tabularnewline
-2 & -2\tabularnewline
\end{tabular}\tabularnewline
 & \tabularnewline
\multicolumn{2}{l}{}\tabularnewline
\multicolumn{2}{>{\raggedright}p{0.5\columnwidth}}{(d)\, FastBCS continues with $C_3^\prime$\\
Clear the tiesets: $tie[1..5] = \emptyset$ \\
Forward step: $i \gets 4$}\tabularnewline
\end{tabular}

\caption{Example: the main steps taken in stage 3 that result in a forward
step to stage 4.}
\label{figureMainSteps3}
\end{figure}

The final stages, as shown in [Figure \ref{figureSecondPass}], follow a series of backward steps from stage 4 to stage 2 where the search ends [25]. The algorithm outputs \emph{pi} and the tau-path statistic calculated from the concordance matrix generated from \emph{pi} [28].

\begin{figure}[ht]
\begin{centering}
\begin{minipage}[c][1\totalheight][t]{1\columnwidth}%
\begin{center}
\begin{minipage}[c][1\totalheight][t]{0.5\columnwidth}%
\textbf{Stage 4f}

Backward Elimination
\begin{quote}
There are no ties to save.
\end{quote}
Tie Logic
\begin{quote}
$pi[4]=4$ is not a member of any tieset for $k>4$.\end{quote}
\end{minipage}\textbf{}%
\begin{tabular}{c|rrrr|r}
\textbf{$C_{4f}$} & \textbf{4} & \textbf{2} & \textbf{1} & \multicolumn{1}{r}{\textbf{5}} & 3\tabularnewline
\hline 
\textbf{4} & 0 & 1 & 1 & -1 & -1\tabularnewline
\textbf{2} & 1 & 0 & -1 & -1 & -1\tabularnewline
\textbf{1} & 1 & -1 & 0 & -1 & -1\tabularnewline
\textbf{5} & -1 & -1 & -1 & 0 & 1\tabularnewline
\cline{2-5} 
3 & -1 & -1 & -1 & \multicolumn{1}{r}{1} & 0\tabularnewline
\hline 
\hline 
$colsums$ & 1 & -1 & -1 & \multicolumn{1}{r}{-3} & \tabularnewline
$ties\_i$ &  &  &  & \multicolumn{1}{r}{} & \tabularnewline
\hline 
\end{tabular}
\par\end{center}

\begin{center}
\begin{minipage}[c][1\totalheight][t]{0.5\columnwidth}%
\textbf{Stage 3f}

Backward Elimination
\begin{quote}
Transpose $pi[3]$ and $pi[2]$.
\end{quote}
Tie Logic
\begin{quote}
$pi[3] = 1$ is not a member of any tieset for $k>3$.
\end{quote}
The $until$ test succeeds. Halt. \\
\\
$pi = [4, 1, 2, 5, 3]$ \\
${\text{\em{tau-path}}} = [1, 1, 0.333, -0.333, -0.4]$ \\%
\end{minipage}\textbf{}%
\begin{tabular}{c|rrrrr}
\textbf{$C_{3f}$} & \textbf{4} & \textbf{2} & \textbf{1} & 5 & 3\tabularnewline
\hline 
\textbf{4} & 0 & 1 & \multicolumn{1}{r|}{1} & -1 & -1\tabularnewline
\textbf{2} & 1 & 0 & \multicolumn{1}{r|}{-1} & -1 & -1\tabularnewline
\textbf{1} & 1 & -1 & \multicolumn{1}{r|}{0} & -1 & -1\tabularnewline
\cline{2-4} 
5 & -1 & -1 & -1 & 0 & 1\tabularnewline
3 & -1 & -1 & -1 & 1 & 0\tabularnewline
\hline 
\hline 
$colsums$ & 2 & 0 & 0 &  & \tabularnewline
$ties\_i$ &  & 2 & 1 &  & \tabularnewline
\hline 
\end{tabular}
\par\end{center}%
\end{minipage}
\par\end{centering}

\caption{Example: the second pass through stages 4f and 3f resulting from a forward step.}\label{figureSecondPass}
\end{figure}

FastBCS guarantees only that a \emph{sequentially maximal} monotone decreasing path is generated (\cite{bib.Yu:2011}, p. 102). It does not lead to a unique permutation nor to a unique tau-path since the choice of ties at earlier stages may limit subsequent maximizations. For example, for the input $X = [1, 2, 3, 5, 4]$ and $Y = [2, 4, 1, 3 , 5]$ in [Figure \ref{figureSecondPass}], our implementation of the algorithm will not find the permutation $pi = [1, 2, 5, 3, 4]$ with a tau-path of $[1, 1, 1, 0.333, 0.2]$. Instead, it finds the permutation $pi = [3, 5, 4, 1, 2]$ with a corresponding tau-path of $[1, 1, 0.333, 0.333, 0.2]$. To find the former tau-path requires a stronger algorithm such as the FullBCS (\cite{bib.Yu:2011}, p.103).

The FastBCS2 algorithm shown in [Algorithm \ref{algorithm-fastbcs2}] is being introduced to reduce the $O(n^3)$   running time of FastBCS to $O(n^2)$   as evidenced by the performance times of FastBCS2 illustrated with [Table \ref{tableRuntimes}] in Subsection \ref{results}. It has the same inputs and outputs as well as overall structure, but some of the operations have been moved, altered, or merged. The differences will be described in Subsection \ref{computationalcomplexity}.

\begin{algorithm}
\caption{FastBCS2($X$, $Y$)}
\label{algorithm-fastbcs2}
\begin{algorithmic}[1]

\Statex \textbf{Input:} A random sample of $n$   pairs $(X_1, Y_1), \ldots , (X_n, Y_n)$   from a continuous bivariate population.
\Statex \textbf{Output:} The final permutation $pi$ and the tau-path statistic derived from the concordance matrix constructed from $pi$.
\Require   $n$ = $length(X) = length(Y)$; $n > 1$   
\Statex // The function $indexOf(a, v)$ returns the index of the value $v$ in array $a$.
\State $C \gets$ the concordance matrix for $(X_1, Y_1), \ldots , (X_n, Y_n)$  
\Statex $pi \gets [1 \ldots n]$; \enspace $tie[1 \ldots n] \gets \emptyset$   
\Statex $i \gets n$  

\For{$j \gets 1 \ldots n$}  
    \State $colsum[j] \gets \sum_{u=1}^n C_{[pi[u], pi[j]]}$   
\EndFor

\Repeat
    \State $ties\_i \gets \{pi[l], l \gets 1 \ldots i | colsum[pi[l]] = minsum \}$
    \If{$|ties\_i| > 1$}
        \State $tie[i] \gets ties\_i$
         \State $r\_tie \gets$ randomly select a member of $ties\_i$
         \State $l \gets indexOf(pi, r\_tie)$
         \State transpose $pi[i]$ and $pi[l]$; transpose $colsum[i]$ and $colsum[l]$
    \Else
        \State $l \gets indexOf(pi, ties\_i[1])$
        \State transpose $pi[i]$ and $pi[l]$; transpose $colsum[i]$ and $colsum[l]$
    \EndIf
        
    \For{$j \gets 1 \ldots i$}
            \State $colsum[j] \gets colsum[j] - C_{[ \pi(i),\pi(j)]}$
    \EndFor

    \For {$k \gets n$ \textbf{downto} $(i+1)$}   
        \If {$pi[i] \in tie[k]$}
            \State $qi \gets [\sum_{u=1}^i C_{[pi[u],pi[i]]}, \sum_{u=1}^{i+1} C_{[pi[u],pi[i]]}, \ldots ,  \sum_{u=1}^k C_{[pi[u],pi[i]]}]$
            \State $qk \gets [\sum_{u=1}^i C_{[pi[u],pi[k]]}, \sum_{u=1}^{i+1} C_{[pi[u],pi[k]]}, \ldots , \sum_{u=1}^k C_{[pi[u],pi[k]]}]$
            \If {$All(qk[u] \geq qi[u])$\ and\ $Any(qk[u] > qi[u]), u=1 \ldots k-i+1$}
                \For{$j \gets i \ldots k$}
                        \For{$l \gets 1 \ldots j$}
                            \State $colsum[l] \gets colsum[l] + C_{[pi(j),pi(l)]}$
                        \EndFor
                    \EndFor
                
                \State transpose $pi[i]$ and $pi[k]$; transpose $colsum[i]$ and $colsum[l]$
                \State $i \gets k - 1$
                \State $tie[m] \gets [], \forall m \leq k$
                \For{$j \gets 1 \ldots i$}
                    \State $colsum[j] \gets colsum[j] - C_{[pi(i),pi(k)]},$
                \EndFor  
                \State GOTO \textbf{repeat}
            \EndIf
            \EndIf
        \EndFor

    \State  $i \gets i - 1$
\Until{$\left(\sum_{j=1}^i \sum_{k=1}^i C_{[pi[k],pi[j]])}) = (i * (i - 1)\right)$}

\State $k \gets i$
\State $T[2] \gets \cdots \gets T[k] \gets 1$
\State
\Return The final permutation $pi$ and the tau-path statistic \{$T[2] \ldots T[n]$\}   

\end{algorithmic}
\end{algorithm}

\subsection{Computational Complexity}
\label{computationalcomplexity}

The analysis of an algorithm requires computational models of the target platforms that will be used for execution of the implementations. Our analysis of FastBCS was based on a random-access machine (RAM) model where operations are assumed to be performed strictly in sequence in constant time, and a data parallel model which defines computation as a sequence of instructions or \emph{kernel} applied synchronously to sets of \emph{processing elements} each with access to private data memory. The programming model for parallelism will be discussed in more detail in Subsection \ref{parallelismmethods} in the context of performance data collected from multicore and manycore devices. No attempts were made to model memory architectures, although the hierarchies and distribution of memory play an important role in the runtime cost of implementations of the FastBCS* algorithms.

\subsubsection{Analysis of FastBCS}
\label{analysisoffastbcs}

Our analysis of FastBCS focused on the runtime as determined by the cost and frequency of critical operations and inner loops. To simplify the analysis, the FastBCS algorithm was rewritten as procedural pseudocode with a syntax resembling familiar programming languages such as C, Pascal or Java (\cite{bib.Cormen:2009}, pp. 20--22). [Table \ref{tableFrequencyAnalysis}] shows the frequency analysis and order of growth of critical statements of the \emph{repeat} loop of the algorithm. The baseline implementation of FastBCS in Java was instrumented to generate experimental data for validating order of growth using the doubling ratio (\cite{bib.Sedgewick:2011}, pp. 192--193) and as a basis for the probabilistic analysis of the average-case running time. Probes were placed in the FastBCS source code just before lines 3, 8, 13, 17, 20, and 24 of the algorithm. We profiled the execution of an implementation of FastBCS using 1,000 samples for each of   $n$ = 200 to 2,000 by 200 to determine the frequency of events at the probes, and to collect statistics on the forward distance, the position of \emph{k} when a forward step was taken, the tieset size, and the value of \emph{i} when the halting condition was met. The empirical cost models derived from this profile are also shown in [Table \ref{tableFrequencyAnalysis}].

\begin{table}[ht!]
\centering{
\setlength{\tabcolsep}{0.10em}
\def\arraystretch{0.8}%
\begin{tabular}{|c|l|c|c|}
\hline 
Line & Statement & \ensuremath{\sim} Frequency & Big-Oh \tabularnewline
\hline 
2 & repeat &  & \tabularnewline
\hline 
3 & \quad for $j \leftarrow 1 \ldots i$ do &  & \tabularnewline
\cline{1-1} \cline{3-4} 
4 & \qquad for $u \leftarrow 1 \ldots i$ do &  & \tabularnewline
\cline{1-1} \cline{3-4} 
4a & \qquad \quad $colsum[j] \leftarrow C_{[pi[u],pi[j]]}$ & $E[X_R] * (i^2)$ & $O(n^3)$\tabularnewline
\hline 
5 & \quad $minsum \leftarrow$ min value in colsum &  & \tabularnewline
\hline 
6 & \quad for $i \leftarrow 1 \ldots i$ do &  & \tabularnewline
\cline{1-1} \cline{3-4} 
6a & \qquad if $(colsum[(pi[i]] = minsum)$ &  & \tabularnewline
\cline{1-1} \cline{3-4} 
6b & \qquad \quad add $pi[i]$ to ties\_i & $E[X_R] * \sum_{k=1}^i(k)$ & $O(n^2)$\tabularnewline
\hline 
7 & \quad if $(ties\_i > 1)$ &  & \tabularnewline
\cline{1-1} \cline{3-4} 
8 & \qquad $tie[i] \leftarrow ties\_i$  &  & \tabularnewline
\cline{1-1} \cline{3-4} 
9 & \qquad $r\_tie \leftarrow$ randomly select an observation &  & \tabularnewline
\cline{1-1} \cline{3-4} 
10 & \qquad $l \leftarrow indexOf(pi, r\_tie)$ & $E[X_R]/2 * n/2$ & $O(n^2)$\tabularnewline
\cline{1-1} \cline{3-4} 
11 & \qquad transpose $pi[i]$ and $pi[l]$ &  & \tabularnewline
\hline 
12 & \quad else &  & \tabularnewline
\cline{1-1} \cline{3-4} 
13 & \qquad $l \leftarrow indexOf(pi, ties\_i[1])$  & $E[X_R]/2 * n/2$ & $O(n^2)$\tabularnewline
\cline{1-1} \cline{3-4} 
14 & \qquad transpose $pi[i]$ and $pi[l]$ &  & \tabularnewline
\hline 
15 & \quad for $k \leftarrow n$ downto $(i+1)$ do &  & \tabularnewline
\cline{1-1} \cline{3-4} 
16 & \qquad if $pi[i]$ member $tie[k]$ & $E[X_R] * \sum_{k=1}^{n-2}(k)$ & $O(n^2)$\tabularnewline
\cline{1-1} \cline{3-4} 
17 & \qquad \quad $qi \leftarrow cumsum(C, pi[i], pi[i], pi[k])$, \ldots &  & \tabularnewline
\cline{1-1} \cline{3-4} 
18 & \qquad \quad $qk \leftarrow cumsum(C, pi[k], pi[i], pi[k])$, \ldots &  & \tabularnewline
\cline{1-1} \cline{3-4} 
19 & \qquad \quad if $All(qk >= qi)$ and $Any(qk >= qi)$ &  & \tabularnewline
\cline{1-1} \cline{3-4} 
20 & \qquad \qquad transpose $pi[i]$ and $pi[k]$ &  & \tabularnewline
\cline{1-1} \cline{3-4} 
21 & \qquad \qquad $i \leftarrow k - 1$ &  & \tabularnewline
\cline{1-1} \cline{3-4} 
22 & \qquad \qquad  for $1 \ldots k$ do &  & \tabularnewline
\cline{1-1} \cline{3-4} 
22a & \qquad \qquad \quad $tie[m] \leftarrow [ ]$ &  & \tabularnewline
\cline{1-1} \cline{3-4} 
23 & \qquad \qquad  GOTO repeat &  & \tabularnewline
\hline 
24 & \quad $i \leftarrow i - 1$ &  & \tabularnewline
\hline 
25 & \quad until $\sum_{j=1}^i \sum_{k=1}^i C_{[pi[k],pi[j]]} = (i * (i-1))$ & $E[X_R] * (i^2)$ & $O(n^3)$\tabularnewline
\hline 
\end{tabular}

\begin{tabular}{ll}
 & \tabularnewline
Empirical Cost Models & Expected value of: \tabularnewline
\hline 
$E[X_R] = -22.06 + 0.97 * n$ & number of iterations in repeat loop\tabularnewline
$E[X_T] = 0.494 + 0.00001 * n$ & number of ties found\tabularnewline
$E[X_M] = -0.08 + 0.19 * n$ & number of times observation is member of tieset\tabularnewline
$E[X_{FD}] = 0.51 + 0.001 * n$ & the total forward distance\tabularnewline
$E[X_H] = 21.61 + 0.03 * n$ & value of \textit{i} when the algorithm halts \tabularnewline
\hline 
\end{tabular}
}
\caption{Frequency analysis of FastBCS.} \label{tableFrequencyAnalysis}
\end{table}

We begin by examining the frequency of the outermost loop. Recall that the FastBCS algorithm divides into two major sections within the \emph{repeat} loop [2--24]. Backward elimination [3--14] performs the search for the observation to eliminate from the subset at each stage \emph{i}. The tie logic [15--23] looks forward, beginning at the end of the tau-path being constructed, to determine if another member of a tieset created in a previous stage \emph{k} that includes the observation $pi[i]$ might improve the Kendall's tau coefficient for concordance matrix $ C_i $ at stage \emph{i}. 

The backward step [24] is taken unless a forward step [21] is reached. While each backward step results in the execution of only one stage, each forward step increases the number of stages to reconsider by $k - i - 1$, a value we call the \emph{forward distance}. The frequency of the \emph{repeat} loop is
$ E[X_R] = E[X_B] + E[X_{FD}] - E[X_H]$, where $E[X_B]$ is the expected number of backward steps, $E[X_{FD}]$ is the expected forward distance and $E[X_H]$ is the expected value of \emph{i} when the algorithm terminates.

The number of backward steps is at most $n - 1$, since the $repeat$ loop will surely terminate [25] when $i = 1$.
 Whether a forward step is taken depends on the number and size of tiesets created, on whether the observation at stage \emph{i} is a member of a tieset generated in a previous stage [7, 16] and on whether a correction to the tau-path under construction needs to be made [19]. A forward step erases all tiesets up to stage \emph{k} [22], and restarts the backward search at $ k-1 $ [23].

\begin{table}[ht]
\centering{
\begin{tabular}{r|r@{\extracolsep{0pt}.}lr|c|c|c|c}
$n$ & \multicolumn{3}{c}{$N_R$ (line 3)} & $N_T$ (line 8) & Tieset size & $i_H$ & $N_T/N_R$\tabularnewline
 & \multicolumn{2}{c}{(avg)} & (max) & (avg) & (avg) & (avg) & \tabularnewline
\hline 
200 & 177&44 & 199 & 86.92 & 3.6 & 22.30 & 0.49\tabularnewline
400 & 367&22 & 394 & 182.59 & 3.8 & 32.71 & 0.50\tabularnewline
600 & 559&76 & 589 & 280.69 & 3.9 & 40.74 & 0.50\tabularnewline
800 & 753&51 & 794 & 380.03 & 4.1 & 47.41 & 0.50\tabularnewline
1000 & 948&04 & 1028 & 480.20 & 4.1 & 53.38 & 0.51\tabularnewline
1200 & 1142&79 & 1209 & 580.73 & 4.2 & 58.89 & 0.51\tabularnewline
1400 & 1338&15 & 1410 & 680.78 & 4.3 & 63.66 & 0.51\tabularnewline
1600 & 1533&73 & 1601 & 782.30 & 4.3 & 68.15 & 0.51\tabularnewline
1800 & 1729&84 & 1788 & 883.69 & 4.4 & 72.81 & 0.51\tabularnewline
2000 & 1925&58 & 2008 & 986.21 & 4.4 & 76.75 & 0.51\tabularnewline
\end{tabular}
}
\caption{Statistics of the count of iterations of the $repeat$ loop ($N_R$),  the count of the times ties were found ($N_T$), the size of a tieset (Tieset size), the value of $i$ when the algorithm halted ($X_H$), and the proportion of times ties were found within the $repeat$ loop ($N_T/N_R$).} \label{tableCountIterations}
\end{table}

Since backward elimination is performed every iteration and summing the columns [3--4a] dominates the order of growth with   $O(n^3)$, it was important to understand the contribution of backward and forward steps to the frequency of the outermost loop. Statistics for the frequency of the loop and tiesets are shown in [Table \ref{tableCountIterations}]. The number of forward steps is negligible. Empirically it was at most 4 for $n \leq 2,000$ (Column $N_{FS}$ in [Table \ref{tableExecutionContexts}]). Since the average number of iterations (Column $N_R$) is less than \emph{N}, we assumed $E[X_R] \simeq n$. Although tiesets are generated about half of the time (Column $ N_T/N_R$), they are small in size (Column ``Tieset size''), decreasing the likelihood that an observation associated with stage \emph{i} will be a member of a previous tieset [17].

\begin{table}[ht]
\centering{
\def\arraystretch{0.8}%
\setlength{\tabcolsep}{0.17em}
\begin{tabular}{r|c | c | c | c | c | c | c}
$n$ & $N_R$ (line 3) & $N_{M}$(line 17) & \multicolumn{2}{c}{$N_{FS}$(line 20)} & \multicolumn{2}{c|}{$N_{FD}$} & $N_{M}/N_R$\tabularnewline
& (avg) & (avg) & (avg) & (max) & (avg) & (max) & \tabularnewline
\hline 
200 & 177.44 & 39.72 & 0.13 & 3 & 0.7 & 21 & 0.22\tabularnewline
400 & 367.22 & 77.37 & 0.13 & 2 & 0.9 & 26 & 0.21\tabularnewline
600& 559.76 & 117.30 & 0.18 & 2 & 1.5 & 31 & 0.21\tabularnewline
800 & 753.51 & 157.26 & 0.17 & 3 & 1.9 & 41 & 0.21\tabularnewline
1000 & 948.04 & 196.40 & 0.20 & 4 & 2.4 & 81 & 0.21\tabularnewline
1200 & 1142.79 & 235.18 & 0.21 & 3 & 2.7 & 64 & 0.21\tabularnewline
1400 & 1338.15 & 274.71 & 0.22 & 3 & 2.8 & 73 & 0.21\tabularnewline
1600 & 1533.73 & 315.88 & 0.20 & 3 & 2.9 & 71 & 0.21\tabularnewline
1800 & 1729.84 & 352.78 & 0.24 & 3 & 3.7 & 59 & 0.20\tabularnewline
2000 & 1925.58 & 391.29 & 0.18 & 3 & 3.3 & 89 & 0.20\tabularnewline
\end{tabular}
}
\caption{Statistics of the count of the iterations of the $repeat$ loop ($N_R$), the number of times the observation of stage $i$ was a member of a previously saved tieset ($N_M$), the number of forward steps ($N_{FS}$), the total forward distance ($N_{FD}$), and the proportion of times tie membership was found to the number of iterations of the $repeat$ loop ($N_M/N_R$).} \label{tableCountIterations2}
\end{table}

\begin{figure}[ht]
\includegraphics[width=4.5in]{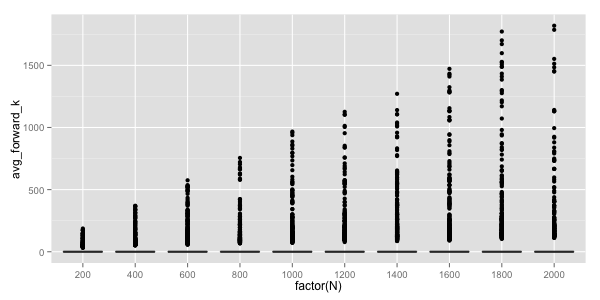}
\caption{The distribution of the average $k$th reset position in a forward step.}\label{figureResetDistribution}
\end{figure}

\begin{figure}[ht]
\includegraphics[width=4.5in]{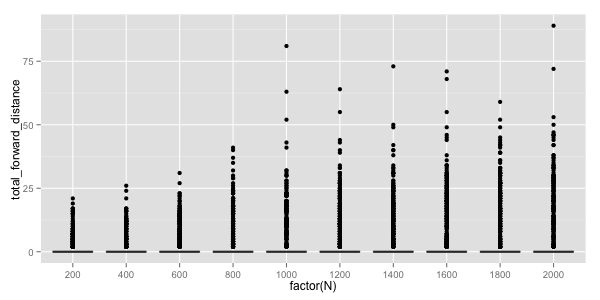}
\caption{The distribution of the total forward distance.}\label{figureForwardDistance}
\end{figure}

\begin{figure}[ht]
\includegraphics[width=4.5in]{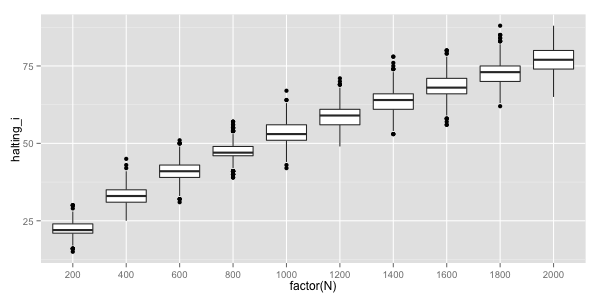}
\caption{The distribution of the expected value of $i$ when the algorithm halts.}\label{figureExpectedValue}
\end{figure}

In [Table \ref{tableCountIterations2}] approximately 20\% of observations at stage \emph{i} are members of a previous tieset (Column $ N_M/N_R$). The distribution of the reset position of \emph{k} for those samples in which a forward step was reached is shown in the boxplot of [Figure \ref{figureResetDistribution}]. The cost for repeating stages resulting from a higher \emph{k} as compared with a lower $k$ increases quadratically. Similarly, the distribution of the total forward distance for these samples is shown in [Figure \ref{figureForwardDistance}]. The increase in the number of iterations by forward steps is offset by early termination when permutations result in fully concordant matrices before $i = 1$. The distributions of \emph{i} at termination for different $n$ are shown in [Figure \ref{figureExpectedValue}].

To reduce the order of growth, we chose the operations that contributed most to the runtime cost and could be adapted to a parallel environment. The computation of column sums [3--4a] in FastBCS became the focus for FastBCS2, and the platform for exploring how to improve the runtime cost of TKTP in a variety of computing environments. The strategies of implementation will be described in Section \ref{efficiencyandimplementationstrategies}, but we first provide a brief overview of the changes that were made to FastBCS.

\subsubsection{Analysis of FastBCS2}
\label{analysisoffastbcs2}

The FastBCS2 algorithm is shown in [Algorithm \ref{algorithm-fastbcs2}]. The central idea of FastBCS2 was to minimize the calculations required to produce the column sums from the permuted concordance matrix $C_i$ at each stage \emph{i}. The summing of all columns of the initial concordance matrix \emph{C} is done once with a cost of $O(n^2)$  before the \emph{repeat} loop is entered and the results are saved in the array \emph{colsum} [2--3]. 

Once the loop is entered, the \emph{colsum} values requiring changes are updated whenever a transposition of observations occurs through the permuted index \emph{pi} [10, 13, 15, 24, 27--28] or a forward step is taken [21--23]. How the incremental adjustments upward or downward are made to the column sums [14--15, 21--23, 27--28] can be seen in [Figure \ref{figureColSum}]. The naturally ordered concordance matrix \emph{C} is created and the column sums in [Figure \ref{figureColSum}(a)] are initialized as $ \{-2, -2, -2, 0, -2\} $. After the observation 5 is transposed with observation 1 [10], the values of the row labeled 1 are subtracted from the column sums shown in [Figure \ref{figureColSum}(a)] leaving the values of the column sums shown in [Figure \ref{figureColSum}(b)]. The adjustment [15] is made only 5 times. Only the operation that sums the columns in a forward step [21--23] can be $O(n^3)$   in the worst-case but, as we have seen, forward steps are rarely taken. The others [14--15, 27--28] are $O(n^2)$ for the RAM computational model.

\begin{figure}[ht]
\begin{minipage}[c][1\totalheight][t]{1\columnwidth}%
\begin{center}
\begin{minipage}[c][1\totalheight][t]{0.5\columnwidth}%
\textbf{(a)} The initialization of \emph{colsums} for C\emph{.}%
\end{minipage}\textbf{}%
\hfill
\begin{tabular}{c|rrrrr|r}
\textbf{$C$} & \textbf{1} & \textbf{2} & \textbf{3} & \textbf{4} & \multicolumn{1}{r}{\textbf{5}} & \tabularnewline
\cline{1-6} 
\textbf{1} & 0 & -1 & -1 & 1 & -1 & \tabularnewline
\textbf{2} & -1 & 0 & -1 & 1 & -1 & \tabularnewline
\textbf{3} & -1 & -1 & 0 & -1 & 1 & \tabularnewline
\textbf{4} & 1 & 1 & -1 & 0 & -1 & \tabularnewline
\textbf{5} & -1 & -1 & 1 & -1 & 0 & \tabularnewline
\cline{1-6} 
colsums & -2 & -2 & -2 & 0 & \multicolumn{1}{r}{-2} & $\gets$\tabularnewline
\cline{1-6} 
\end{tabular}
\par\end{center}

\begin{center}
\begin{minipage}[c][1\totalheight][t]{0.5\columnwidth}%
\textbf{(b)} The $colsums$ for $C_5$ after transposing $pi[5]$ and $pi[1]$ in stage 5. %
\end{minipage}\textbf{}%
\hfill
\begin{tabular}{c|rrrrr|r}
\textbf{$C_{5}$} & \textbf{5} & \textbf{2} & \textbf{3} & \textbf{4} & \multicolumn{1}{r}{\textbf{1}} & \tabularnewline
\cline{1-6} 
\textbf{5} & 0 & -1 & 1 & -1 & -1 & \tabularnewline
\textbf{2} & -1 & 0 & -1 & 1 & -1 & \tabularnewline
\textbf{3} & 1 & -1 & 0 & -1 & -1 & \tabularnewline
\textbf{4} & -1 & 1 & -1 & 0 & 1 & \tabularnewline
\textbf{1} & -1 & -1 & -1 & 1 & 0 & $\gets$\tabularnewline
\cline{1-6} 
colsums & -1 & -1 & -1 & -1 & \multicolumn{1}{r}{-2} & $\gets$\tabularnewline
\cline{1-6} 
\end{tabular}
\par\end{center}%
\end{minipage}

\caption{An example of how FastBCS2 distributes the column sum calculations.
(a) The initial values for the column sums are calculated before the
\emph{repeat} loop {[}Line 4{]}. 25 operations are performed. (b)
The values of the row $pi[5] = 1$ are subtracted from the \emph{colsums}
of \emph{C} to produce the \emph{colsums} of  $C_5$ {[}Lines 14-15{]}.
5 operations are performed.}\label{figureColSum}
\end{figure}

Refactoring and distributing the column sum operations has two advantages. First, it allows the postponement of operations until they are needed. Second, it provides an opportunity for an implementation to use source code fragments as specifications for parallelism for implementations using compilers that can translate them for data-parallel computing environments. In the example in [Figure \ref{figureColSum}], were $n \leq 1,000$ and the implementation running on an accelerated processing unit (APU) with 1,000 processing elements (PE), the \emph{for} loop could theoretically be done in constant time by assigning each $ PE_i $ the subtraction operation and the \emph{i}th elements of the column sum array of $C_5$ and the transposed row $pi[5] = 1$. In practice, it is not straightforward. The next section discusses why. 

\subsection{Efficiency and Implementation Strategies}
\label{efficiencyandimplementationstrategies}

Three implementations of the FastBCS and FastBCS2 algorithms were developed. The two of FastBCS in R (\cite{bib.RCoreTeam:2015}) and Java were used to explore and cross-validate the algorithm's implementations. FastBCS2, the third implementation, was an optimization of FastBCS implemented in Java and parameterized to execute critical sections in either sequential or sequential-parallel mode. This subsection provides a brief history of the implementations of the algorithms, the opportunities for optimization provided by parallelism, our methods and results, and a discussion of the results.

\subsubsection{Early Implementations}
\label{earlyimplementations}

The first reference implementation of the FastBCS algorithm was done in the R programming language to explore the tau-path algorithm. Optimization was not a goal. However, the need to screen larger datasets required more efficient implementations. Although the R environment is mostly implemented in C, type safety checking imposes a performance penalty when data is passed between R and C. To avoid this, the second implementation of FastBCS was written entirely in C. Using variables of size   $n = 512$, our benchmarks showed a nearly 30x improvement in runtime performance over the R implementation. However, we found the C implementation unable to handle large toxicity microarray datasets and difficult to profile. To address these limitations and to begin exploring the parallelism made possible by the introduction of Java 8, we created a faithful reproduction in Java of the original R implementation which became the baseline for runtime performance analysis.

\subsubsection{The Promise of Parallelism}
\label{thepromiseofparallelism}

There are four parallel architecture categories in Flynn's taxonomy (\cite{bib.Rauber:2013}, p. 11): Single-Instruction, Single-Data (SISD); Multiple-Instruction, Multiple-Data (MISD); Single-Instruction, Multiple-Data (SIMD); and Multiple-Instruction, Multiple-Data (MIMD). Of these, the SIMD (manycore) and MIMD (multicore) architectures were of interest. In SIMD architectures, all processing elements execute the same instruction synchronously at each step, but each processing element has access to private data memory. In MIMD architectures, each processing element has private access to program and data memory, and each element works independently. For applications with a high degree of data parallelism, the SIMD approach can be very efficient and---excluding the broader context of integrating with software applications---simpler to program than MIMD computers. This is because a single program flow controls execution groups of processing elements, eliminating the need for synchronization at the program level (\cite{bib.Rauber:2013}, p. 11) Recent advances in hardware and software created opportunities to reduce the order of growth of TKTP by integrating algorithm design with implementations for data-parallel computational models. 

Graphics Processing Units (GPUs) now contain thousands of processing elements. Where GPUs were once devoted to rendering pipelines for realtime graphics applications, their value to other data-intensive applications is now recognized. Heterogeneous computing environments are emerging which combine aspects of MIMD and SIMD architectures with a unified memory model. A unified memory eliminates the need to transfer large blocks of data between the hierarchies of memory devoted to either CPU or GPU. 

While algorithms often cannot be easily parallelized, many contain islands of computation that would be amenable to parallel execution with the right kind of programming language and system software support. Three things are required. First is a parallel programming model that provides an abstraction of a heterogeneous computing environment to insulate the software developer from the low-level software and hardware differences of each device. Second, the programming language used for the implementation must allow sequential and parallel fragments to be expressed uniformly and in a declarative form that allows late-binding decisions to be made by the virtual machine as to which available compute device would most efficiently run various elements of the computation. Third, because software engineering tools are critical for algorithm design and profiling of the implementation, standards are needed in order for these tools to be produced. Software technology and standards have advanced far enough to make it possible to explore implementing algorithms as sequential-parallel high-level programs as we have done with FastBCS2* implementations.

\subsubsection{Methods and Contexts}
\label{parallelismmethods}

The OpenCL standard (OpenCL) computing model views a system as a collection of compute devices such as CPUs or GPUs, each of which may contain multiple processing elements. OpenCL 1.0, based on the C99 specification, was a language for describing program fragments called \emph{kernels} that execute on compute devices. Although we considered C++\slash OpenCL 1.0 for specifying kernels, we chose Java. In Java version 8, Oracle introduced parallel streams, lambda expressions, and language extensions that provide a declarative style of programming for parallelism. An experimental Java library (\cite{bib.Aparapi}) was also being developed at AMD to provide support for GPU parallelization. OpenCL, Java 8, and Aparapi became the foundation for our initial work on sequential-parallel algorithms.

The first attempt to parallelize FastBCS was based on the Aparapi library. The sequential-parallel implementation was specified in Java. At runtime in the Java virtual machine (JVM), the GPU-parallel fragments compiled into Java bytecode were converted by Aparapi into the instructions and dispatched to the computing system's GPU device. For the software developer, this eliminated the need to write and debug the OpenCL code that managed the dispatching of work items into queues. Our initial attempts at speedup were disappointing. Because of updates to the critical data structures for each stage of FastBCS2, the cost of data transfers and Aparapi's overhead were significant and offset the gains from GPU computation. We refocused our efforts on what could be achieved using the extensions to Java 8 in multicore systems.

The gains from the redesign of FastBCS2 came by moving the initial calculation of column sums outside the scope of the \emph{repeat} loop and distributing only the updates across the algorithm. The FastBCS2 gains come from the redesign of the FastBCS algorithm, reducing the order of growth in the average case from $O(n^3)$ to $O(n^2)$   as evidenced in [Table \ref{tableRuntimes}] of Subsection \ref{results}. This also made possible the use of the parallel streams framework in Java 8. The FastBCS2 implementation is parameterized to execute either sequentially or with parallel fragments on multicore CPUs. In sequential-parallel mode, the parallel streams are executed. The creation and initialization of the concordance matrix, column summation, and column sum updates attempt to run on as many cores as are available. 

To separate the effects of algorithm design, programming language, computational platform, and parallelism on runtime performance, the data were collected in six contexts on three platforms as described in [Table \ref{tableExecutionContexts}]. A \emph{context} consisted of an implementation of either the FastBCS or FastBCS2 algorithm (FB or FB2); the programming language used to author the implementation and the runtime environment or virtual machine in which it was executed (R or Java); the hardware and operating system (P1, P2, or P3); and an execution mode as either sequential (s) or sequential-parallel (sp).

\begin{table}[ht]
\centering{
\def\arraystretch{0.8}%
\setlength{\tabcolsep}{0.25em}
\begin{tabular}{|cc|cccc|ccc|}
\cline{3-9} 
\multicolumn{1}{c}{} &  & \multicolumn{4}{c|}{Software} & \multicolumn{3}{c|}{Hardware}\tabularnewline
\hline 
Context & Algorithm & Lang. & SDK & OS & Mode & Device & Cores & Mem.\tabularnewline
\hline 
FB-RP1s & FastBCS & R & RStudio & OSX & s & MP-I7 & 4 & 16GB\tabularnewline
FB-JP2s & FastBCS & Java & Java 8 & OSX  & s & MP-I7 & 4 & 16GB\tabularnewline
FB2-JP2s & FastBCS2 & Java & Java 8 & OSX & s & MP-I7 & 4 & 16GB\tabularnewline
FB2-JP3s & FastBCS2 & Java & Java 8 & Linux & s & EC2-Xeon & 16 & 32GB\tabularnewline
FB2-JP2sp & FastBCS2 & Java & Java 8 & OSX & sp & MP-I7 & 4 & 16GB\tabularnewline
FB2-JP3sp & FastBCS2 & Java & Java 8 & Linux & sp & EC2-Xeon & 16 & 32GB\tabularnewline
\hline 
\end{tabular}
}
\caption{The six execution contexts. \textbf{Context:} The contexts are encoded as algorithm (FB or FB2),
programming language and virtual machine  (R or Java), the computing platform (P1, P2, or P3), and the execution mode ("s" or "sp"). \textbf{Algorithm:} The algorithm being implemented. \textbf{Lang:} The statistical language R, or Java.  \textbf{SDK:} The Software Development Kit used to load, or compile and run the implementations was RStudio 0.99.441 for R and the Oracle  Java Development Kit 1.8.0\_b25-17, respectively. \textbf{OS:} The operating system was either OSX 10.11 or Linux Ubuntu 14.04.2.LTS.
\textbf{Mode:} The execution mode indicates whether the implementation was run sequentially (s) or as sequential-parallel (sp). \textbf{Device:} The Apple computer was
a Macbook Pro with a 2.2 Ghz Intel I7 (MP-I7). The larger multi-core
system was an Amazon EC2 c4.8xlarge instance with a 2.9 Ghz Intel Xeon E5-2666v3
(EC2-Xeon). \textbf{Cores:} The number of cores in the microprocessor.
\textbf{Mem:} The size (GB) of random access memory available to the CPUs.}\label{tableExecutionContexts}
\end{table}

\begin{figure}[ht] 
\centering{
\includegraphics[scale=0.65]{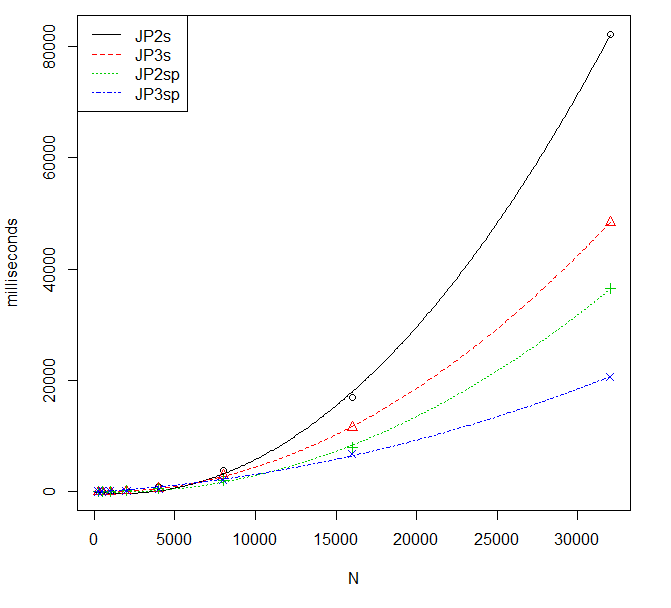}}
\caption{Quadratic fits to average runtimes of FastBCS2.}
\label{FigQuadraticFits}
\end{figure}  

\subsubsection{Results}
\label{results}
Three performance metrics were generated for each context: runtime, speed-up, and the doubling ratio. The runtime measures the total amount of work performed by all processors. The input for each implementation was a pair of variables $X$ and $Y$ from a uniform distribution with input sizes that doubled from $n = 250$ to 32,000. The execution of an implementation of the FastBCS* algorithm was performed five times for each pair and the average time in milliseconds recorded. No other applications were running on the platform during the test. The results are given in [Table \ref{tableRuntimes}] and quadratic fits to average runtimes depicted in [Figure \ref{FigQuadraticFits}], with relative performance given in [Table \ref{tableSpeedupComparisons}]. Due to memory and time constraints, we were unable to complete the R implementation of FastBCS (FB-RP1s) for $n > 4,000$.   

\begin{table}[ht]
\centering{
\def\arraystretch{0.8}%
\setlength{\tabcolsep}{0.20em}
\begin{tabular}{lrrrrrrrr}
\cline{2-9} 
 & \multicolumn{8}{c}{Vector length ($n$)}\tabularnewline
Context & 250 & 500 & 1,000 & 2,000 & 4,000 & 8,000 & 16,000 & 32,000\tabularnewline
\hline 
FB-RP1s & 441 & 3,358 & 42,189 & 341,974 & 1,325,750 & {*} & {*} & {*}\tabularnewline
FB-JP2s & 17 & 82 & 575 & 4,685 & 34,850 & 303,481 & 2,368,399 & 19,709,329\tabularnewline
FB2-JP2s & 5 & 10 & 58 & 214 & 916 & 3,876 & 16,867 & 82,237\tabularnewline
FB2-JP3s & 3 & 10 & 33 & 129 & 579 & 2,922 & 11,586 & 48,431\tabularnewline
FB2-JP2sp & 31 & 33 & 65 & 171 & 564 & 2,083 & 7,929 & 36,506\tabularnewline
FB2-JP3sp & 38 & 38 & 83 & 212 & 560 & 2,167 & 6,733 & 20,636\tabularnewline
\hline 
\end{tabular}}
\caption{A table of the runtimes of different implementations of FastBCS* for pairs of variables $X$ and $Y$ of size $n = 200$ to 2,000 by 200 generated from a uniform distribution. Each cell is the average runtime in milliseconds of five executions. Cells marked with an asterisk were not completed because of memory and time constraints on the platform used for RStudio.}\label{tableRuntimes}
\end{table}

The speedups shown in [Table \ref{tableSpeedupComparisons2}] were derived from the runtime values. \emph{Speedup} is defined as $S(n) = T^*(n) / T(n)$, where $T^*(n)$ is the execution time of a reference context and $T(n)$ the execution time of the context being compared to it. The use of a sequential implementation as a reference is especially helpful in determining the benefit of parallelism (\cite{bib.Rauber:2013}, p. 162). The Java implementation of FastBCS (FB-JP1s) was used as a reference for comparing speedup across all contexts. Three additional tables using different references for each comparison are shown in [Table \ref{tableSpeedupComparisons2}] to illustrate the effect on speedup by the redesign of the algorithm [Table \ref{tableSpeedupComparisons2}(a)], the number of cores in multicore systems [Table \ref{tableSpeedupComparisons2}(b)], and the benefits of the Java 8 parallel stream framework [Table \ref{tableSpeedupComparisons2}(c)].

\begin{table}[ht]
\centering{
\def\arraystretch{0.8}%
\setlength{\tabcolsep}{0.30em}
\begin{tabular}{lrrrrrrrr}
\cline{2-9} 
 & \multicolumn{8}{c}{Vector length ($n$)}  \tabularnewline
Context & 250 & 500 & 1,000 & 2,000 & 4,000 & 8,000 & 16,000 & 32,000 \tabularnewline
\hline 
FB-RP1s & 0.038 & 0.024 & 0.013 & 0.012 & 0.009 & {*} & {*} & {*}\tabularnewline
FB-JP2s & 1.0 & 1.0 & 1.0 & 1.0 & 1.0 & 1.0 & 1.0 & 1.0\tabularnewline
FB2-JP2s & 3.2 & 8.2 & 9.9 & 21.9 & 38.0 & 78.3 & 140.4 & 239.7\tabularnewline
FB2-JP3s & 5.3 & 7.9 & 17.4 & 36.2 & 60.2 & 103.9 & 204.4 & 407.0\tabularnewline
FB2-JP2sp & 0.5 & 2.5 & 8.9 & 27.4 & 61.8 & 145.7 & 298.7 & 539.9\tabularnewline
FB2-JP3sp & 0.4 & 2.1 & 6.9 & 22.1 & 62.3 & 140.1 & 351.8 & 955.1\tabularnewline
\hline 
\end{tabular}
}
\caption{Speedup comparing all contexts using FB-JP2s as the reference implementation. Cells marked with an asterisk could not be computed because the runtime values were not generated.}\label{tableSpeedupComparisons}
\end{table}

\begin{table}[ht]
\centering{
\setlength{\tabcolsep}{0.30em}
\begin{tabular}{llrrrrrrrr}
\cline{3-10} 
 &  & \multicolumn{8}{c}{Vector length ($n$)}   \tabularnewline
 & Context & 250 & 500 & 1,000 & 2,000 & 4,000 & 8,000 & 16,000 & 32,000 \tabularnewline
\hline 
\hline 
(a) & FB-JP2s & 1.0 & 1.0 & 1.0 & 1.0 & 1.0 & 1.0 & 1.0 & 1.0\tabularnewline
 & FB2-JP2s & 3.2 & 8.2 & 9.9 & 21.9 & 38.0 & 78.3 & 140.4 & 239.7\tabularnewline
\hline 
\hline 
(b) & FB2-JP2s & 1.0 & 1.0 & 1.0 & 1.0 & 1.0 & 1.0 & 1.0 & 1.0\tabularnewline
 & FB2-JP3s & 1.6 & 1.0 & 1.8 & 1.7 & 1.6 & 1.3 & 1.5 & 1.7\tabularnewline
\cline{2-10} 
 & FB2-JP2sp & 1.0 & 1.0 & 1.0 & 1.0 & 1.0 & 1.0 & 1.0 & 1.0\tabularnewline
 & FB2-JP3sp & 0.8 & 0.9 & 0.8 & 0.8 & 1.0 & 1.0 & 1.2 & 1.8\tabularnewline
\hline 
\hline 
(c) & FB2-JP2s & 1.0 & 1.0 & 1.0 & 1.0 & 1.0 & 1.0 & 1.0 & 1.0\tabularnewline
 & FB2-JP2sp & 0.2 & 0.3 & 0.9 & 1.3 & 1.6 & 1.9 & 2.1 & 2.3\tabularnewline
\cline{2-10} 
 & FB2-JP3s & 1.0 & 1.0 & 1.0 & 1.0 & 1.0 & 1.0 & 1.0 & 1.0\tabularnewline
 & FB2-JP3sp & 0.1 & 0.3 & 0.4 & 0.6 & 1.0 & 1.3 & 1.7 & 2.3\tabularnewline
\hline 
\end{tabular}
}
\caption{A table of speedup comparisons using different references to show: (a) the effect on speedup of
the redesigned algorithm FastBCS2 compared with FastBCS; (b) the effect
on speedup of a 16-core system as compared with a 4-core system for
sequential or sequential-parallel execution as well as factors such as processor clock speeds, hardware architecture, and cache sizes and speeds; and (c) the effect on speedup
of sequential execution as compared with sequential-parallel execution
on the same platform for two different platforms.}\label{tableSpeedupComparisons2}
\end{table}

The doubling ratio was generated using an implementation based on the algorithm shown in [Algorithm \ref{algorithm-doubling-ratio}]. It provides an empirical measurement for approximating most common orders of growth. The doubling hypothesis is that as the limit of the ratio approaches \ensuremath{\sim}$2^b$, the running time is approximately $an^b$ and this is an acceptable approximation of $O(n^b)$ when the ratio is being used to predict performance (\cite{bib.Sedgewick:2011}, p. 193). The results of the doubling ratio across all platforms are shown in [Table \ref{table:binaryLogOfDoublingRatio}].

\begin{algorithm}[ht]
\caption{CalculateDoublingRatios($n_A, n_B$) } 
\label{algorithm-doubling-ratio}
\begin{algorithmic}[1]

\Statex \textbf{Input:} $n_A$ and $n_B$   are the lower and upper limits of $n$.  
 
    \State $ITERATIONS \gets 5$
    \State $i \gets 1$
    \State $n \gets n_A$  
    \State $fnc \gets$ A function that invokes an implementation of FastBCS*
    \State $T_i \gets timeFastBCS (fnc, n, ITERATIONS)$
    \While{$n \leq n_B$}  
        \State $i \gets i+1$
        \State $n \gets 2n$
        \State $T_i \gets timeFastBCS (fnc, n, ITERATIONS)$  
        \State $ratios[i] \gets T_i/T_{i-1}$
        \If{$ratios[i]$ approaches a limit}
            \State \textbf{break}
        \EndIf
    \EndWhile
    \State \Return $ratios$
    \Statex
    \Function {timeFastBCS}{$fnc$, $n$, $ITERATIONS$} 
        \State $t \gets 0$
        \For{$i \gets 1 \ldots iterations$}
            \State Generate variables $X$ and $Y$ of size $n$ from a uniform distribution.
            \State $t \gets t + runtime(fnc(X, Y))$
        \EndFor
        \State \Return $t / iterations$
    \EndFunction        

\end{algorithmic}
\end{algorithm}

\begin{table}[ht]
\centering{
\def\arraystretch{0.8}
\begin{tabular}{lrrrrrrrr}
\cline{2-9} 
 & \multicolumn{8}{c}{Vector length ($n$)}  \tabularnewline
Context & 250 & 500 & 1,000 & 2,000 & 4,000 & 8,000 & 16,000 & 32,000 \tabularnewline
\hline 
FB-RP1s & NA & 2.9 & 3.7 & 3.1 & 3.2 & {*} & {*} & {*} \tabularnewline
FB-JP2s & NA & 2.3 & 2.8 & 3.0 & 2.9 & 3.1 & 3.0 & 3.1\tabularnewline
FB2-JP2s & NA & 0.9 & 2.5 & 1.9 & 2.1 & 2.1 & 2.1 & 2.3\tabularnewline
FB2-JP3s & NA & 1.7 & 1.7 & 2.0 & 2.2 & 2.3 & 2.0 & 2.1\tabularnewline
FB2-JP2sp & NA & 0.1 & 1.0 & 1.4 & 1.7 & 1.9 & 1.9 & 2.2\tabularnewline
FB2-JP3sp & NA & 0.0 & 1.1 & 1.4 & 1.4 & 2.0 & 1.6 & 1.6\tabularnewline
\hline 
\end{tabular}
}
\caption{The binary logarithm of the doubling ratio, or $lg(T(2n)/T(n))$, where 
$T$ is the execution time of an implementation of FastBCS{*} for $X$ and
$Y$ from a uniform distribution. The values of the cells represent the
power $b=lg(ratio)$ which is an acceptable approximation of the order of growth   $O(n^b)$
when used to predict performance using the doubling hypothesis. Cells
marked with an asterisk could not be computed because the runtime values were not generated.}\label{table:binaryLogOfDoublingRatio}
\end{table}

\subsubsection{Discussion}
\label{discussion}

The first two rows of [Tables \ref{tableRuntimes} and \ref{tableSpeedupComparisons}] show dramatic improvements in speedup in large $n$ ($n=32,000$) from just two changes: 1) the use of a static language and its development and runtime environments (Java, SDK, and JVM) over a dynamic language (R and RStudio) and 2) minor changes to the algorithm's design based on a frequency analysis and an empirical study of the algorithm's behavior resulting from a randomization of the inputs. Comparing the doubling ratios of FB-RP1s with FB-JP2s and FB-JP2s with FB2-JP2s at   $n=$ 2,000 and 4,000, shows that most of the improvement to the order of growth comes from the algorithm's redesign. 

Further improvements come from parallelization through the use of the Java 8 parallel streams for larger   $n$. We see this by comparing FB2-JP2s with FB2-JP2sp and FB2-JP3s with FB2-JP3sp. In [Tables \ref{tableSpeedupComparisons2}c and \ref{table:binaryLogOfDoublingRatio}] speedup and the order of growth begin to improve on the 4-core processor at   $n=2,000$, and on the 16-core processor at   $n=8,000$. More powerful hardware improves the speedup for a sequential implementation (rows FB2-JP2s and FB2-JP3s of [Table \ref{tableSpeedupComparisons2}b]) but has no apparent effect on the order of growth (rows FB2-JP2s and FB2-JP3s for $n \geq 2,000$ in [Table \ref{table:binaryLogOfDoublingRatio}]).

Since the FastBCS* algorithms operate mostly on homogeneous numeric data and a significant proportion of execution time in FastBCS* implementations is spent on the calculation of concordance matrix column sums, manycore systems theoretically could further reduce the order of growth by mapping a computational kernel onto the data of a column or group of columns, allowing parallel calculation of each column's sum. In our initial work with GPUs, however, we found that data transfer cost was a limitation with the current generation of GPUs. When the GPU does not share memory with the host CPU, data transfers are required to maintain a coherent state between the CPU and GPU memories. The data transfer costs overwhelm any gain from parallelism. The next generation of heterogeneous computing environments will eliminate the need for data transfers through the inclusion of unified memory among devices.

Standards for heterogeneous computing environments are emerging and the major object-oriented languages such as Java as well as newer languages such as Dart or Swift have the language features needed to express data-parallelism. Although advancements in heterogeneous computing environments make parallel programming increasingly attractive for algorithms like TKTP, the tools for developing sequential-parallel implementation to run heterogeneously have not yet reached the maturity needed to allow these methods to be more broadly applied. We expect this to change within the next couple of years. Until then, implementations for sequential-parallel algorithms written in high level languages can be designed and executed on current generation multicore CPUs.


\section{Power and Accuracy Simulation Study} \label{PowerAccuracy}

In this section, simulation studies are used to exemplify the performance of TKTP in screening for sample points that come from a population in which the variables are associated.  Since the method is based only on the relative rankings of values observed for each variable, it is invariant to scale transformations of the variables, and thus it is sufficient to limit the investigation of properties to copulae, which have uniform margins.  Samples are simulated under three different types of mixing with the independent copula: 1) Frank, 2) Gaussian, and 3) a positively associated Frank copula and a negatively associated Gaussian copula.  The objectives are to identify a suitable measure of performance of TKTP in screening for subsamples that support monotonic association of the variables; to check the robustness of this measure under different distributions with comparable association; and to provide guidelines for setting the operational values of $\alpha$ and $w$ to meet coverage requirements.  Toward these aims, it is helpful to provide some pertinent background on the Frank and Gaussian copula families. For a thorough introduction, see \cite{bib.Nelsen:2006}.

\begin{figure}[ht] 
\centering{
\includegraphics[scale=0.65]{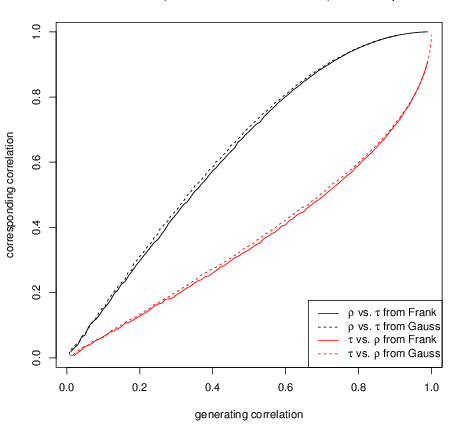}} 
\caption{Corresponding population values of Kendall's $\tau$ and Spearman's $\rho$ from Frank and Gaussian copulae.}
\label{Fig4.1}
\end{figure}

Although the Frank copulae are often indexed by $\tau$ and the Gaussian copulae indexed by the population version $\rho$ of Spearman's correlation coefficient, which is also the product moment correlation of variables $X$ and $Y$ after their individual cumulative distribution function (CDF) transforms, [Figure \ref{Fig4.1}] shows how either parameter may be used to index either family of copulae. In particular the Frank copulae with $\tau = 0.3, 0.5 {\text { or }} 0.7$ have corresponding values of $\rho = 0.45, 0.7 {\text{ or }} 0.89$, respectively.

Each line in [Figure \ref{Fig4.1}] was formed from 100 summary points, with each point based on a sample of 100,000 points from one of the copulae.  For each sample generated from a Frank copula with a given $\tau$, a Spearman correlation coefficient $\rho$ was calculated, which essentially equals the population $\rho$ since the sample size is so large.  Similarly, samples from the Gaussian copulae were generated with fixed values of $\rho$ and the corresponding population $\tau$ was estimated from these large samples.  Note that for any copula the population value of Spearman's $\rho$ is the same as the population (Pearson) product moment correlation between the uniformly distributed marginal variables.

In the figure, the solid black line is based on points generated from Frank copulae, and the solid red line is based on points generated from Gaussian copulae.  Dashed lines are inversions of the solid ones, and colors indicate either $\rho$ versus $\tau$ (black) or $\tau$ versus $\rho$ (red).  Although there exist distributions in which the relationship between population $\tau$ and $\rho$ may be quite different from that given above, this relationship remains quite stable across the Frank and Gaussian copulae, as seen from the slight difference between solid and dashed lines of the same color.

\begin{figure}[ht] 
\centering{
\includegraphics[scale=0.9]{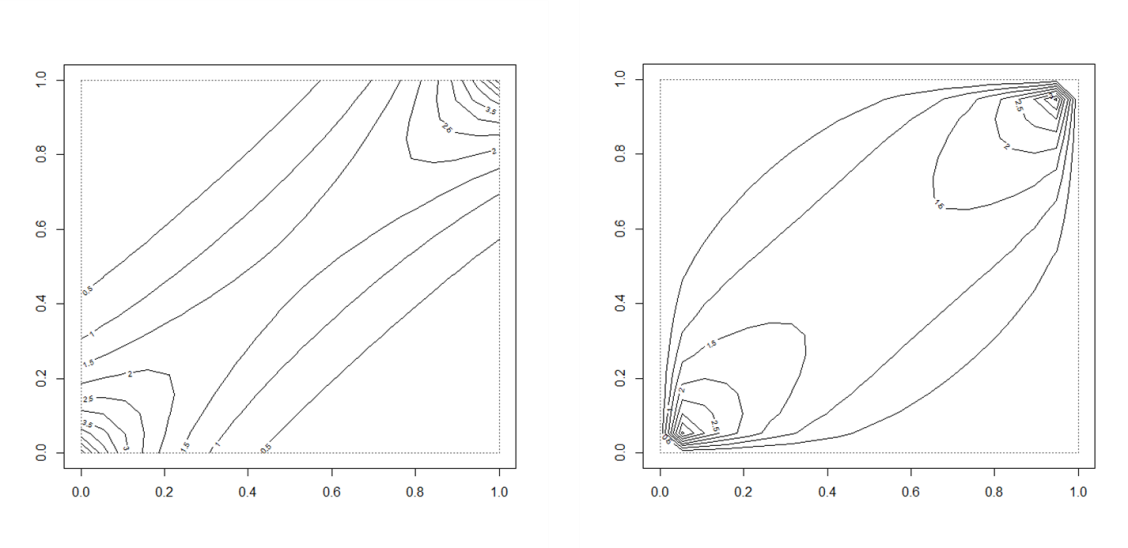}} 
\caption{Density contours of Frank (left) and Gaussian (right) copulae, both with $\tau$  = 0.5 and $\rho$ = 0.7. Contours depict density levels of 0.5 to 5 in steps of 0.05.}
\label{Fig4.2}
\end{figure}

Although a Frank and a Gaussian copula may share the same values of $\tau$ and $\rho$, the distributions themselves are generally quite distinct.  Density contours of the Frank and Gaussian copulae, both having $\tau = 0.5$ and $\rho = 0.7$, are depicted in [Figure \ref{Fig4.2}].

\begin{figure}[ht] 
\centering{
\includegraphics[scale=0.6]{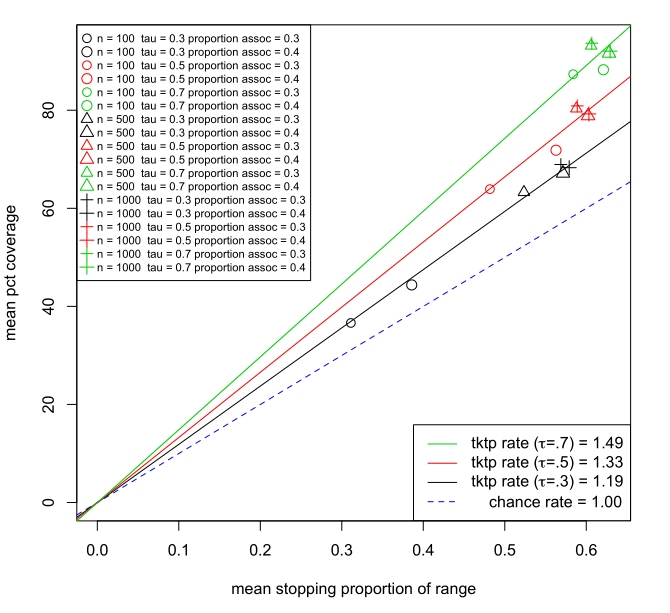}} 
\caption{Summary of TKTP ($\alpha = 0.05, w = 3$) simulations under Frank mixtures of copulae.}
\label{Fig4.3}
\end{figure}

Now consider the performance of TKTP when applied first to various mixtures of Frank and independent copulae.  This first simulation study follows a 3 $\times$ 3 $\times$ 2 complete block design, with three levels of sample sizes $(n = 100, 500, 1000)$, three levels of association strength in the subsample $(\tau = 0.3,$ $0.5,$ $0.7)$ and two levels of mixing proportion of the associated subsample $(p = 0.3, 0.4)$.  For each of the 18 combinations of these parameters, 100,000 samples are generated from the Frank mixture, and the results of the $(\alpha = 0.05; w = 3)$ TKTP screen recorded. [Figure \ref{Fig4.3}] summarizes the major finding.

In [Figure \ref{Fig4.3}], each point represents the paired means from 100,000 simulations. ``Rate'' refers to the increase in percent coverage with increase in percent of sample selected.  The plot character represents sample size (circle $\rightarrow n$ =  100; triangle $\rightarrow n$ =  500; plus $\rightarrow n$ = 1,000).  Color represents strength of association (black $\rightarrow \tau$ = 0.3; red $\rightarrow \tau$ = 0.5; green $\rightarrow \tau$ = 0.7). The size of each symbol represents the proportion of associated subsample in the mix (small $\rightarrow$ 30\% of observed pairs are associated; large $\rightarrow$ 40\%).

The stopping point is the point at which the algorithm determines the predictive power of the remaining sample reduces to chance levels.  Percent covered is the percent of observations from the Frank copula that appear before the stopping point and are thus included in the screen.

Increasing sample size from $n = 100$ to $n = 500$ has a big effect on both stopping point and percent coverage, but differences between these variables at $n = 500$ and $n = 1,000$ are small.  The biggest effect of sample size is on the distributions of these two variables, which are highly skewed left for small sample sizes, but approach normality for larger sample sizes.

A larger underlying proportion (0.4 versus 0.3) always increases the mean stopping proportion of the range for any sample size and any $\tau$, but does not necessarily increase mean coverage probability.  The method has little power to detect proportions less than $2\sqrt{n} - 1.758 n^{1/6}$, which is the expected length of the longest increasing subsequence under independence (\cite{bib.Baik:1999}, \cite{bib.Aldous:1999}), and it is also difficult for the algorithm to distinguish a large associated subpopulation from the whole population.  In the limited range investigated here, size of the underlying associated subpopulation does not have a big effect on the algorithm.

On the other hand, strength of association in the subpopulation has a substantial effect on the performance of TKTP.  Stronger association (increasing $\tau$) produces a substantial increase in both mean stopping point and mean coverage for any fixed sample size, which is captured by the rate of coverage, as given by the slopes of the lines in [Figure \ref{Fig4.3}]. Formally, under any given copula,
\begin{equation}
\text{Rate of coverage} = \frac{\text{Mean(Number of associated values selected)}} {\text{Mean(Stopping point)}}, \label{RateOfCoverage}
\end{equation}
which clearly does not depend on the sample size.  [Table \ref{tableRateOfCoverage}] gives the rates of coverage under Frank and Gaussian copulae that are matched for the same strength of association in the subsample.

\begin{table}[ht]
\centering
\begin{tabular}{| c | c | c | c |}
\hline
$\tau$ & Frank Rate & Gaussian Rate & $\rho$\\
\hline
0.3 & 1.19 & 1.17 & 0.45\\
0.5 & 1.33 & 1.30 & 0.70\\
0.7 & 1.49 & 1.46 & 0.89\\
\hline
\end{tabular}
\caption{Rates of Coverage [Equation (\ref{RateOfCoverage})] under Frank and Gaussian copulas.} \label{tableRateOfCoverage}
\end{table}

The simulations compare coverage rates for TKTP for $\alpha$ in $\{0.1, 0.05, 0.10\}$ to see how stable the coverage probability is in this range. It is enough to consider just Frank copulae with $n = 500$, mixing proportion = 0.3, and $\tau$ in $\{0.3, 0.5, 0.7\}$. The standard errors of these rates are all about 0.1, suggesting little real difference in the TKTP performance; the comparable Gaussian version of [Figure \ref{Fig4.3}] is virtually indistinguishable from that figure, and is thus omitted.  Nevertheless, the actual coverage is systematically less in all 18 Gaussian copulae under study than in their comparable Frank copulae, suggesting that the small differences indicate a real but slight underperformance of TKTP in the Gaussian setting.

In all cases, the rate of coverage seems to capture the effectiveness of the method.  From the table, it remains stable under different types of mixtures; and, from [Figure \ref{Fig4.3}], it is not affected by small changes in the mixing proportion.  

The control parameters $\alpha$ and $w$ of the TKTP have different tasks.  The value $\alpha$ changes the distribution of the stopping point, with smaller values of $\alpha$ inducing stochastically larger distributions of the stopping point.  The effect is much the same as selecting an operating point in a classification method. The value $\alpha = 0.05$ gives a rough midrange value, and lowering it slightly will increase both the stopping point and the coverage probability, but lowering it too far will greatly increase the probability of wrongly including uncorrelated samples in the selection set.  For large sample sizes $n \ge 500$, we recommend $\alpha = 0.05$ because these quantiles are well estimated in the simulations and the stopping point does not change much for other values of $\alpha$ near 0.05. The control parameter $w$ helps to stabilize the performance of TKTP.  It is useful mostly for small sample sizes $n \ge 100$, as studied here; we recommend $w = 3$, the smallest practical smoothing.

\section{Application: Long Term Predictive Power of Oil for S\&P 500 Stocks} \label{Application}
As an example of how TKTP may be used in large datasets, the weekly closing prices of stocks currently listed in the Standard \& Poor's 500 index and having a 10-year history are correlated against the corresponding price of oil six months prior. By eliminating weeks where the overall pattern is broken, the TKTP method provides not only a robust estimate of correlation, but also the set of common time points over which predictions are most reliable for a key cluster of stocks.  For most of the stocks in this cluster, oil is not a direct cause of the performance of these stocks, but simply a reliable indicator of the likely course of stock prices, at least during the time periods ascertained by the TKTP. The goal here is not to predict actual stock prices, but to identify likely periods of a sustained trend.

\begin{figure}[H]
\centering{
\includegraphics[scale=0.55]{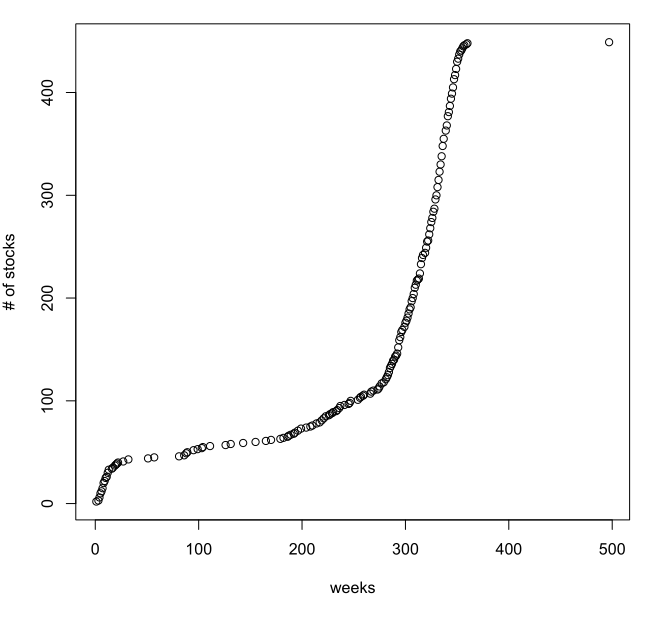}} 
\caption{Cumulative number of stocks associated with 6-month lagged oil.}
\label{Fig5.1}
\end{figure}

The dataset is composed of weekly prices of 26-week lagged S\&P 500 stocks versus oil over a decade from 12/31/2004 to 1/2/2015 (523 time points, 449 stocks with 10 years of data).  The initial screen follows a two-step selection process:

Step 1.  Find stocks with partial association over the most weeks. There were $523-26 = 497$ time point pairs (oil time, stock time) available. The first screen was to select stocks associated over at least 60\% of these time point pairs. There were 273 such stocks correlated with 6-month oil over periods of at least 6 years.  See [Figure \ref{Fig5.1}] which suggests this cutoff.

Step 2.  Find the correlations (both Pearson and Kendall) between oil and lagged stock prices over the TKTP-selected time point pairs. There are 77 of these with Pearson correlations over 0.9, listed in [Table \ref{tableStockOilCorrelation1}].

\begin{table}[ht]
\centering
\begin{adjustbox}{width=1\textwidth}
\begin{tabular}{| l | c | c | c | c | c | c | c | c | c | r |}
\hline
FMC &   COL &  TJX  &  CNP &  BEN  &  XOM  &  SIAL &   PH &  NBL &  XRAY &  EMC  \\
0.901 & 0.902 & 0.902 & 0.902 & 0.902 & 0.902 & 0.902 & 0.902 & 0.903 & 0.903 & 0.905 \\
\hline
ISRG & APH & PCAR & DHR & RRC & FTI & FAST & ROK & DRI & EL & RAI \\
0.905 & 0.905 & 0.906 & 0.906 & 0.907 & 0.907 & 0.907 & 0.908 & 0.908 & 0.908 & 0.908 \\
\hline
PWR & SPG & PCLN & LH & CL & AMZN & HCP & UTX & ROST & BLK & IBM \\
0.909 & 0.909 & 0.910 & 0.910 & 0.910 & 0.911 & 0.911 & 0.912 & 0.913 & 0.913 & 0.913 \\
\hline
EW & SO & GWW & KO & PCP & GIS & ALTR & BBBY & CERN & INTU & DTV \\
0.914 & 0.914 & 0.916 & 0.916 & 0.916 & 0.916 & 0.916 & 0.917 & 0.917 & 0.917 & 0.917 \\
\hline
FLS & MCD & XEL & WAT & CTXS & SWK & WEC & ES & AZO & CSX & CMI \\
0.918 & 0.918 & 0.918 & 0.918 & 0.918 & 0.919 & 0.919 & 0.920 & 0.920 & 0.921 & 0.922 \\
\hline
FDO & PSA & DLTR & ORCL & YUM & FFIV & RL & ACN & ED & O & EMR \\
0.922 & 0.922 & 0.923 & 0.923 & 0.924 & 0.924 & 0.924 & 0.926 & 0.926 & 0.926 & 0.928 \\
\hline
VTR & PX & ESRX & AAPL & RHT & VAR & SRCL & CTSH & ARG & CRM & PRGO \\
0.928 & 0.928 & 0.930 & 0.930 & 0.932 & 0.932 & 0.933 & 0.934 & 0.935 & 0.936 & 0.942 \\
\hline
\end{tabular}
\end{adjustbox}
\caption{Pearson correlations for 77 S\&P 500 stocks correlated with 6-month prior oil price over different restricted periods of at least 6-year duration.}
\label{tableStockOilCorrelation1}
\end{table}

\begin{figure}[H]
\centering{
\includegraphics[scale=0.65]{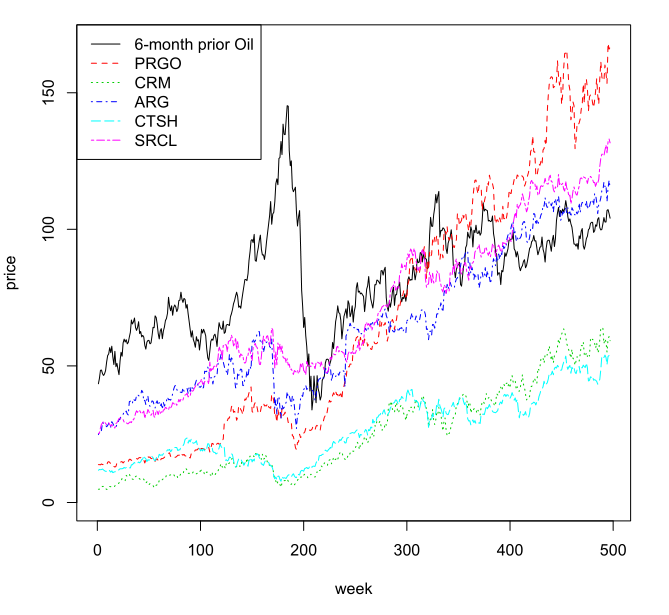}} 
\caption{Five stock prices associated with 6-month lagged oil: 2005-2014.}
\label{Fig5.3}
\end{figure}

[Figure \ref{Fig5.3}] depicts the full 10-year time courses of the weekly price of the five stocks with the largest Pearson correlations in [Table \ref{tableStockOilCorrelation1}] relative to the 6-month prior price of oil.  These five stocks come from different industries: PRGO--pharmaceuticals, CRM--software, ARG--chemicals, CTSH--IT services, SRCL--Commercial Services. This suggests that oil price predicts, at least in pattern, an aspect of the general economy.  Although the huge crash of oil in 2008 was disproportionate, it did predict a downturn in each of the stocks six months later.

More importantly than general association, the TKTP method helps to identify the time periods over which association is strongest.  Pooling information from several stocks strengthens the degree of confidence about the time periods selected, but the pooling needs to be done methodically.  The idea is to pool information from stocks that agree strongly with each other.

A complete-linkage clustering approach is used to find the stocks from which to pool time-point inclusions.  The agreement measure used for pairs of time-point sets is the Jaccard/Tanimoto coefficient $J$ defined for sets $A$ and $B$ as
\begin{eqnarray*}
J  = \frac{|A \cap B|}{|A \cup B|}.
\end{eqnarray*}

A cluster of 24 of the 273 Step 1-screened stocks has all pairs with $J > 0.8$.  These 24 stocks, listed in [Table \ref{tableStockOilCorrelation2}], include three of the five stocks from [Figure \ref{Fig5.3}], and 13 of the stocks from [Table \ref{tableStockOilCorrelation1}]. 

\begin{table}[ht]
\centering
\begin{tabular}{| l | c | c | c | c | r |}
\hline
CAN & ARG & ABC & APH & AZO & BLL\\
HSIC & DTV & ECL & EL & FIS & FISV\\
GIS & HIL & REGN & ROP & CRM & SIAL\\
SRCL & TJX & UNP & UHS & WAT & WEC\\
\hline
\end{tabular}
\caption{Cluster of stocks completely linked by $J > 0.8$.}
\label{tableStockOilCorrelation2}
\end{table}

\begin{figure}[ht] 
\centering{
\includegraphics[scale=0.65]{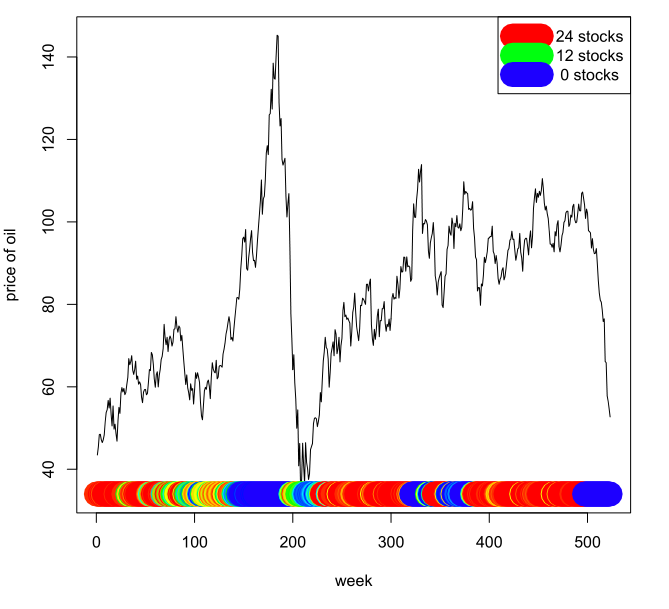}}
\caption{TKTP-selected weeks for oil price prediction of stock cluster prices.}
\label{Fig5.5}
\end{figure}  

The average 24 TKTP correlations with 6-month oil were 0.89 (Pearson) and 0.75 (Kendall). [Figure \ref{Fig5.5}] illustrates, for each week, the number of stocks having that week included by TKTP.  The color coding follows the rainbow with red indicating inclusion in all 24 stocks; yellow-green, in 13 stocks; and blue in 0 stocks. When plotted under the price of oil, it appears that weeks tend to be excluded during the most volatile periods for oil prices.	

\section{Discussion} \label{sectionDiscussion}
The computational efficiencies attained by the new algorithms enable a whole new approach to discovering key subpopulations from big data. Many pairs of variables may be screened to see if there is a common subset of observations that supports strong association between the pairs, and thus represents a subpopulation supporting a whole network. This could be a subpopulation of cancers supporting a gene-network for chemoresistance, or a subpopulation of time periods supporting an economic network for growth. 

Although the methods here have been presented in the classical view of analyzing a random sample from a population, they also apply to any large dataset. Inference here is internal to the dataset treated as its own population. Identified subsets correspond to a real subpopulation, but identifying the subpopulation externally may not be straightforward. For the stock price example, the aphorism, ``A rising tide lifts all boats,'' seems to apply.  A foretelling rise of oil prices, ignoring brief periods of large volatility, signals a rise in tide, at least in the harbor where the cluster of two dozen stocks is moored.  It is not clear which stocks will remain in the harbor, or what other, more specific, influences will be in force.

The decision to focus on ranks rather than on the original numerical values was motivated both by concerns for inferential robustness and insights from recent mathematics. When moving between the database and external environments, calibration issues often arise, but causal relationships persist {\em {in situ}}. Thus rank associations found within a database subset are very likely to appear in corresponding external subpopulations. Among rank-based methods, the multistage ranking model is very flexible since the number of parameters increases with sample size.

Recent mathematical methods motivate the use of the multistage ranking method. In the special case where all the population parameters $\{\theta_j\}$ are equal, yielding the Mallows model, \cite{bib.Starr:2009} has found a limiting distribution which, under proper scaling, is a Frank copula. More general parameterizations of the multistage model appear to converge to mixtures of Mallows, which covers a very wide range of ranking distributions. If this is true, the multistage model for a pair $(X,Y)$ of variables ranking the $n$ observations will converge to a mixture of Frank copulae.  A ``signal'' in such a mixture of might correspond to a mixture of components with large values of Kendall’s tau, with ``noise,'' if present, being represented by a mixture of nearly independent Frank copulae. Thus, in this conceptual setting, TKTP screens out a nearly independent component.  Doing this on the basis of a single ranking is remarkable; finding additional pairs of variables with the same limiting distribution, as in the example of Section \ref{Application}, adds to the accuracy.  See \cite{bib.Awasthi:2014} and \cite{bib.Meila:2012} for general approaches to mixtures of Mallows models.



\bibliography{Bibliography/FullBibliography}
\end{document}